\documentclass[lettersize,journal]{IEEEtran}
\usepackage{amsmath,amsfonts, amssymb}
\usepackage{algorithmic}
\usepackage{tcolorbox}
\usepackage{algorithm}
\usepackage{array}
\usepackage[caption=false,font=normalsize,labelfont=sf,textfont=sf]{subfig}
\usepackage{textcomp}
\usepackage{stfloats}
\usepackage{url}
\usepackage{verbatim}
\usepackage{graphicx}
\usepackage{nicefrac}
\usepackage{cuted}
\usepackage{trfsigns}
\usepackage{cite}
\usepackage{tikz}
\usepackage{booktabs}
\usepackage{pgfplots}
\usepackage[normalem]{ulem}
\usepackage{soul}

\usepackage{amsthm}
\renewcommand{\qedsymbol}{\ensuremath{\blacksquare}}

\usepackage[makeroom]{cancel}

\usepgfplotslibrary{groupplots,fillbetween}
\usetikzlibrary{arrows.meta,matrix}
\tikzset{>/.tip={Triangle[angle=60:2mm]}}
\usetikzlibrary{external}
\definecolor{darkgruen}{RGB}{33,160,71}

\definecolor{uni_apfelgruen}{cmyk}{.5, 0, 1, 0}
\definecolor{uni_mittelblau}{cmyk}{1, 0.4, 0, 0}
\definecolor{uni_bulletblau}{RGB}{49,99,183}
\definecolor{uni_gelb}{cmyk}{0, 0.1, 1, 0}
\definecolor{uni_rot}{cmyk}{0, 1, 1, 0}
\definecolor{uniblauHell}{RGB}{0,190,255}
\definecolor{uniblauDunkel}{RGB}{0,65,145}
\definecolor{unigrau}{RGB}{51,51,51}
\definecolor{darkgray176}{RGB}{176,176,176}

\definecolor{lavenderplot}{RGB}{191,148,228}
\definecolor{coralplot}{RGB}{255,127,80}
\definecolor{cyanplot}{RGB}{37,219,168}

\definecolor{plotlyblue}{RGB}{99,110,250}
\definecolor{plotlyred}{RGB}{239,85,59}
\definecolor{plotlygreen}{RGB}{0,204,150}
\definecolor{plotlypurple}{RGB}{171,99,250}
\definecolor{plotlyorange}{RGB}{255,161,90}
\definecolor{plotlylightblue}{RGB}{25,211,243}
\definecolor{plotlypink}{RGB}{255,102,146}
\definecolor{plotlylightgreen}{RGB}{182,232,128}
\definecolor{plotlylightpink}{RGB}{255,151,255}
\definecolor{plotlyyellow}{RGB}{254,203,82}

\definecolor{cpred}{RGB}{220,138,120}
\definecolor{cptext}{RGB}{76,79,105}

\newif\ifshowComment
\showCommenttrue

\DeclareRobustCommand{\rchi}{{\mathpalette\irchi\relax}}
\newcommand{\irchi}[2]{\raisebox{\depth}{$#1\chi$}}

\newcommand{\xte}{\ensuremath{x_{\mathrm{terr}}}}
\newcommand{\Xte}{\ensuremath{X_{\mathrm{terr}}}}
\newcommand{\nro}{\ensuremath{n_{\mathrm{rot}}}}
\newcommand{\Nro}{\ensuremath{N_{\mathrm{rot}}}}
\newcommand{\nrr}{\ensuremath{n_{\mathrm{rrn}}}}
\newcommand{\Nrr}{\ensuremath{N_{\mathrm{rrn}}}}
\newcommand{\nxr}{\ensuremath{n_{\mathrm{xrn}}}}
\newcommand{\Nxr}{\ensuremath{N_{\mathrm{xrn}}}}

\def\symbolrate{\ensuremath{R_\mathrm{S}}}
\def\simfreq{\ensuremath{f_\mathrm{sim}}}
\def\lolw{\ensuremath{\Delta\nu}}
\def\lint{\ensuremath{\rchi}}
\def\phox{\ensuremath{\vartheta}}
\def\rxslope{b_{1,k}}
\def\rxslopesq{b^2_{1,k}}
\def\txslope{a_{1,k}}
\def\rxinterception{b_{0,k}}
\def\txinterception{a_{0,k}}

\newcommand{\trf}{\,\laplace\,}

\begin{document}

\title{Equalization-Enhanced Phase Noise: Modeling and DSP-aware Analysis}

\author{Sebastian Jung,~\IEEEmembership{Graduate Student Member,~IEEE,} Tim Janz,~\IEEEmembership{Graduate Student Member,~IEEE,} \\ Vahid Aref, and Stephan ten Brink,~\IEEEmembership{Fellow,~IEEE}
  \thanks{The authors are with the Institute of Telecommunications, University of Stuttgart (e-mail: \{jung,janz,tenbrink\}@inue.uni-stuttgart.de), and Nokia, Stuttgart (e-mail: vahid.aref@nokia.com).}
}

\maketitle

\begin{abstract}
  In coherent optical communication systems the laser phase noise is commonly modeled as a Wiener process.
   We propose a sliding-window based linearization of the phase noise, enabling a novel description. 
   We show that, by stochastically modeling the residual error introduced by this approximation, equalization-enhanced phase noise (EEPN) can be described and decomposed into four different components.
   Furthermore, we analyze the four components separately and provide a stochastical model for each of them.
  This novel model allows to predict the impact of well-known algorithms in coherent digital signal processing (DSP) pipelines such as timing recovery (TR) and carrier phase recovery (CPR) on each of the terms.
  Thus, it enables to approximate the resulting signal affected by EEPN after each of these DSP steps and helps to derive appropriate ways of mitigating such effects.
\end{abstract}


\section{Introduction}
\IEEEPARstart{W}{ith} increasing data rates, and therefore increasing symbol rates, equalization-enhanced phase noise (EEPN) becomes a non-negligible impairment for coherent optical systems.
EEPN stems from the nonlinear interaction between the phase noise of local oscillators at the receiver and transmitter and the electronic dispersion compensation (EDC).
For higher symbol rates, the accumulated dispersion is larger and thus an EDC filter with more memory is required which leads to a larger penalty induced by EEPN.
The impact of EEPN on the receiver's signal-to-noise ratio (SNR) is noticeable for current pluggables and will be even larger for the next generation pluggables with 800G and beyond \cite{Xu2023,Xu2024}.
Hence, a fundamental understanding of EEPN is crucial to develop improved mitigation techniques which are required for next generation optical communication systems.

The effect of EEPN has first been described in \cite{Shieh2008} where it was found to create inter-symbol interference (ISI) and induce a timing jitter \cite{Oda2010}.
In other words, amplitude noise is induced by the the phase noise \cite{Xie2009}.
Further investigations in \cite{Kakkar2015, Kakkar2017,Arnould2019} provided more detailed insights into the mathematical description of the phenomenon.
More recently, the focus has shifted towards mitigation techniques \cite{Martins2024,Jung2024,Abolfathimomtaz2024,Qiu2024} as well as finding improved phase noise models to better reflect real world lasers \cite{Xu2023,Xu2024,Wang2024}.

In this work, we present a phase noise model that enables a novel description of EEPN impacted communication links.
It is based on a linearization as a first step and an estimation of the error as a second step.
The result is a mathematical description that allows to separate EEPN into four impairment terms.
A separation of EEPN in four different components has already been shown in \cite{Kakkar2017}.
However, our model enables a stochastical description of each of these terms by the use of a few assumptions and simplifications.
All of these required steps are shown to be reasonable within this work and yield a good approximation for EEPN impacted signals.

\begin{figure*}[]
  \centering
  \begin{tikzpicture}
  \begin{groupplot}[group style={group size=4 by 1}]
    \nextgroupplot[
      xmin=-1.72,
      xmax=1.72,
      ymin=-1.72,
      ymax=1.72,
      tick align=outside,
      tick pos=left,
      width=.27\textwidth,
      height=.27\textwidth,
      xticklabel style={font=\footnotesize},
      yticklabel style={font=\footnotesize},
      title={Timing error term \xte},
      title style={yshift=-.2cm, font=\footnotesize},
      major tick length=.1cm,
    ]
    \addplot graphics [includegraphics cmd=\pgfimage,xmin=-1.72, xmax=1.72, ymin=-1.72, ymax=1.72] {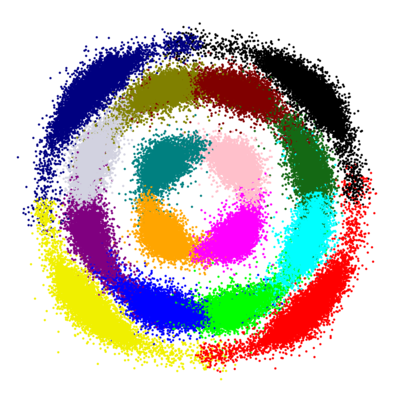};
    \nextgroupplot[
      xmin=-0.29,
      xmax=0.29,
      ymin=-0.29,
      ymax=0.29,
      tick align=outside,
      tick pos=left,
      width=.27\textwidth,
      height=.27\textwidth,
      xticklabel style={font=\footnotesize},
      yticklabel style={font=\footnotesize},
      title={Rotation term \nro},
      title style={yshift=-.2cm, font=\footnotesize},
      major tick length=.1cm,
    ]
    \addplot graphics [includegraphics cmd=\pgfimage,xmin=-0.29, xmax=0.29, ymin=-0.29, ymax=0.29] {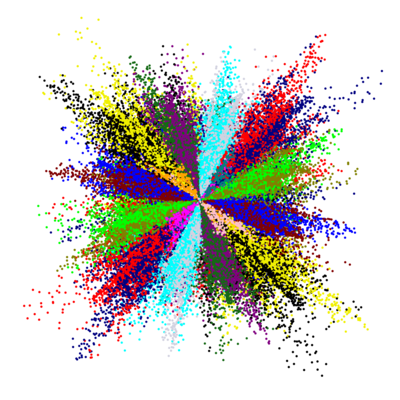};
    \nextgroupplot[
      xmin=-0.3169,
      xmax=0.3169,
      ymin=-0.3169,
      ymax=0.3169,
      tick align=outside,
      tick pos=left,
      width=.27\textwidth,
      height=.27\textwidth,
      xticklabel style={font=\footnotesize},
      yticklabel style={font=\footnotesize},
      title={Receiver residual term \nrr},
      title style={yshift=-.2cm, font=\footnotesize},
      major tick length=.1cm,
    ]
    \addplot graphics [includegraphics cmd=\pgfimage,xmin=-0.3169, xmax=0.3169, ymin=-0.3169, ymax=0.3169] {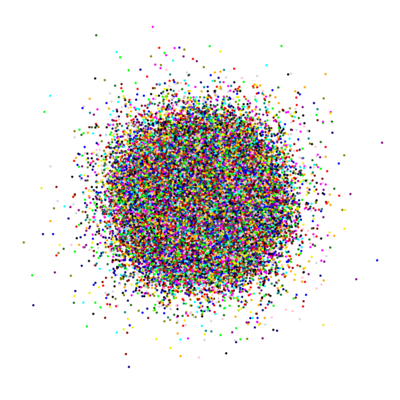};
    \nextgroupplot[
      xmin=-0.0293,
      xmax=0.0293,
      ymin=-0.0293,
      ymax=0.0293,
      tick align=outside,
      tick pos=left,
      width=.27\textwidth,
      height=.27\textwidth,
      xticklabel style={font=\footnotesize},
      yticklabel style={font=\footnotesize},
      title={Cross residual term \nxr},
      title style={yshift=-.2cm, font=\footnotesize},
      major tick length=.1cm,
      scaled x ticks = false,
      scaled y ticks = false,
      xtick = { -0.02, 0 , 0.02},
      xticklabels = { -0.02, 0 , 0.02},
      ytick = { -0.02, 0 , 0.02},
      yticklabels = { -0.02, 0 , 0.02},
    ]
    \addplot graphics [includegraphics cmd=\pgfimage,xmin=-0.0293, xmax=0.0293, ymin=-0.0293, ymax=0.0293] {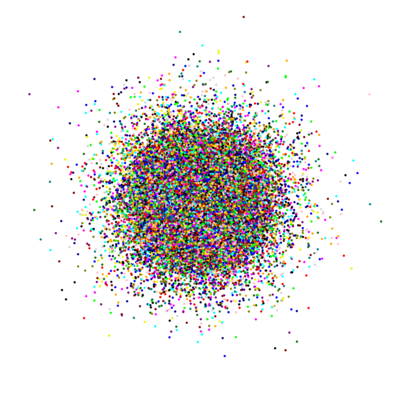};
  \end{groupplot}
\end{tikzpicture}
  \caption{Components of the proposed EEPN model for a 150 kHz linewidth laser. The constellation points of the 16-QAM are color-coded to indicate how they are affected by the respective impairment term, as further detailed in Eq.~(\ref{eq::EEPN_model_time_domain_zero}).}
  \label{fig::EEPN_terms}
\end{figure*}

In Fig.~\ref{fig::EEPN_terms}, the different components of EEPN, as derived by the proposed model, are shown.
The impact per symbol is color coded to visualize the dependency of the respective terms.
A noisy version of the original 16-QAM constellation can be seen in the leftmost subplot, which we call the \emph{timing error term}.
This can be interpreted as the most significant effect of EEPN that has the largest impact on the induced penalty.
Moreover, it also contains the most information about the sent signal.
The second term, called the \emph{rotation term} further rotates the symbols along their phase shift orientation and stems from the consideration of a transmit laser.
This term no longer allows to unambiguously infer the originally sent symbol and the information is lost.
The third and fourth term, i.e., the pure noise terms, which will be referred to as \emph{receiver residual noise term} and \emph{cross residual noise term}, induce a Gaussian error cloud around each symbol.
These terms can be, according to \cite{Kakkar2017}, attributed to be mostly irretrievable as these terms contain low information as they
behave like a white noise.
From these plots it is apparent that the cross residual term has much smaller power than the receiver residual term and is, at the given linewidth, negligible.
These terms will be derived in detail in Section \ref{sec::proposed_phase_noise_model} and \ref{sec::residual_error}.

It is worth mentioning that the proposed model connects the phase noise process directly to its impact on the signal in terms of timing error and additional noise terms  which allows to better characterize the residual error induced by EEPN after each step in a classical coherent digital signal processing (DSP) pipeline.
Moreover, the signal with residual EEPN after timing recovery (TR) and carrier phase recovery (CPR) can be described by our model.
With this, the overall system impact of phase noise can be better characterized which helps developing novel mitigation techniques.
\\\newline
The rest of the paper is organized as follows.
In Section~\ref{sec::related_work}, some recent and relevant contributions about EEPN are briefly reviewed- before the general system model is presented in Section~\ref{sec::system_model}.
Our proposed modification of the phase noise model is presented in Section~\ref{sec::proposed_phase_noise_model}.
The model is refined in Section~\ref{sec::residual_error} yielding a full description of the four terms.
The impact of different DSP algorithms on EEPN penalty with respect to each term is evaluated in Section~\ref{sec::dsp_aware_analysis}.
A quick summary and conclusions are provided in Section~\ref{sec::Conclusion}.

\section{Related Work \label{sec::related_work}}
To the best of out knowledge. a first mathematical description of EEPN was given in \cite{Shieh2008} and numerical simulations were performed to show that the EEPN penalty scales linearly with accumulated chromatic dispersion, i.e., bandwidth and transmission length, as well as the laser linewidth.
Similar simulations in \cite{Xie2009} also showed this interaction between phase noise and EDC, resulting in amplitude noise and significantly degrading the performance in coherent optical communications.

The influence of the phase noise of transmitter and receiver lasers was experimentally studied in \cite{Oda2010}.
It was shown that, with out precompensation of chromatic dispersion, the transmit laser has little impact for link distances over $\sim 100\,\mathrm{km}$.
Additionally, the numerical results for the penalty of EEPN were verified and EEPN was found to induce a timing jitter.

A thorough mathematical description of EEPN was given in \cite{Kakkar2015}.
This descriptions indicate that EEPN is caused by the interference of multiple demodulated and delayed versions of the original signal.
Furthermore, an equation for estimating the SNR penalty induced by EEPN was given and a bandwidth cutoff limit for the mitigation was derived.

In \cite{Kakkar2017}, EEPN was separated into different spectral components.
Each component was then analyzed in terms of its impact on the signal and the possibility to mitigate it.
The components were mainly split into a slow varying mean frequency and a fast and practically untrackable frequency jitter.

Different phase recovery schemes were investigated in \cite{Arnould2019}.
It was shown that even perfect knowledge of the phase noise process does not allow to fully compensate the phase errors.
In addition, it was show that Ideal data remodulation (IDR) with a carefully chosen averaging length can be used in simulation to give a comparable performance to implementable phase recovery algorithms such as blind phase search (BPS) while being much easier to simulate.
Different phase recovery schemes have been investigated in terms of their ability to correctly estimate the EEPN penalty and the memory of EEPN was characterized by means of autocorrelation functions.

A stochastic analysis in \cite{Wang2022} tries to describe the elliptical scatter of received constellation points because of EEPN.s
In \cite{Xu2023}, the authors showed that the phase noise model and real lasers differ in their induced penalty.
Furthermore, a model was presented that correlates the residual timing error caused by EEPN and the SNR penalty.

The residual timing error was shown to be partially mitigatable by classical timing recovery algorithms such as Godard \cite{Godard1978} or Gardner \cite{Gardner1986} in \cite{Martins2024}.
A second-stage timing recovery after a classical coherent DSP  without a specifically tweaked first stage timing error detector was proposed in \cite{Qiu2024} and achieved moderate gains.
We have shown in \cite{Jung2025} how different second-stage timing recovery schemes perform and proposed an optimized version of Mueller-Muller \cite{Mueller1976 } that improved the findings in \cite{Qiu2024}.
More general adaptive filters beyond timing compensation were presented in \cite{Jung2024} and \cite{Abolfathimomtaz2024}.
The latter modeled the phase noise in terms of its Fourier frequencies and derived a filter structure to deal with EEPN for different phase recovery schemes. Both \cite{Jung2024} and \cite{Abolfathimomtaz2024} were able to significantly reduce the EEPN penalty.

Modifying the phase noise model to better fit real world lasers was proposed in \cite{Xu2024} by means of applying filters derived from the power spectral density (PSD) of laser phase noise measurements.
Another modified phase noise model was proposed in \cite{Wang2024} that allows to more accurately predict the EEPN penalty.

From these references, a typical system model is established that is used to investigate EEPN.
The specifics will be presented in the following section.

\section{System model \label{sec::system_model}}
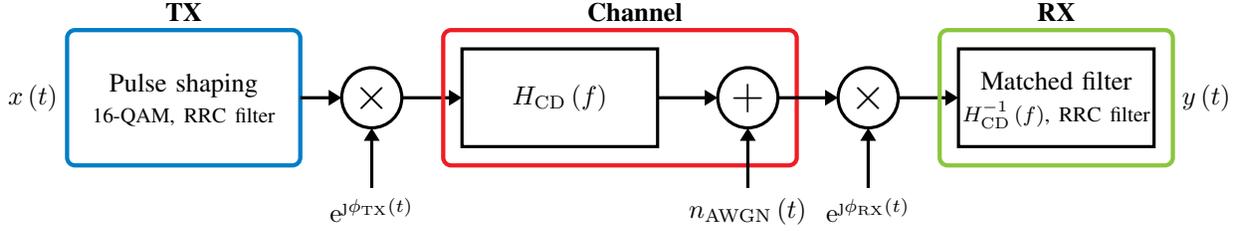
\begin{figure*}
  \centering
  \begin{tikzpicture}

    \def\boxwidth{2.6}
    \def\boxheight{1.3}
    \def\bigboxheight{2}
    \def\roundingrad{3}
    \def\arrowlen{1.2}
    \def\adderrad{.4}
    \def\boundingbox{.25}
    \draw (-\boundingbox, \boxheight/2) node[anchor=east]{\strut $x\left( t \right)$};
    \draw[ultra thick, rounded corners=\roundingrad,color=uni_mittelblau] (-\boundingbox,-\boundingbox) rectangle (\boxwidth+\boundingbox,\boxheight+\boundingbox);
    \draw (\boxwidth/2,\boxheight + \boundingbox/2) node[anchor=south]{\bfseries TX\strut};
    \draw (\boxwidth/2,\boxheight/2) node[text width=80, align=center]{Pulse shaping\strut\\\footnotesize 16-QAM, RRC filter};

    \draw[very thick, ->] (\boxwidth+\boundingbox,\boxheight/2) -- (\boxwidth+\arrowlen-\adderrad,\boxheight/2);
    \draw[very thick](\boxwidth+\arrowlen,\boxheight/2) node[font=\LARGE]{$\times$} circle (\adderrad);
    \draw[very thick,->] (\boxwidth+\boundingbox+\arrowlen-\boundingbox,\boxheight/2-\arrowlen)  node[anchor=north]{$\mathrm{e}^{\j\phi_\mathrm{TX}\left(t\right)}$\strut} -- (\boxwidth+\boundingbox+\arrowlen-\boundingbox,\boxheight/2-\adderrad);

    \draw[very thick, ->] (\boxwidth+\arrowlen+\adderrad,\boxheight/2) -- (\boxwidth+\arrowlen+\arrowlen,\boxheight/2);
    \draw[ultra thick, rounded corners=\roundingrad,color=uni_rot] (\boxwidth+\arrowlen+\arrowlen-\boundingbox,-\boundingbox) rectangle (\boxwidth+\arrowlen+\arrowlen+\boxwidth+\arrowlen+\adderrad+\boundingbox,\boxheight+\boundingbox);
    \draw (\boxwidth+\arrowlen+\arrowlen+\boxwidth/2+\arrowlen/2+\adderrad,\boxheight+\boundingbox/2) node[anchor=south]{\bfseries\strut Channel};
    \draw[very thick] (\boxwidth+\arrowlen+\arrowlen,0) rectangle (\boxwidth+\arrowlen+\arrowlen+\boxwidth,\boxheight);
    \draw (\boxwidth+\arrowlen+\arrowlen+\boxwidth/2,\boxheight/2) node[]{$H_\mathrm{CD}\left( f \right)$\strut};

    \draw[very thick,->] (\boxwidth+\arrowlen+\arrowlen+\boxwidth,\boxheight/2) -- (\boxwidth+\arrowlen+\arrowlen+\boxwidth+\arrowlen-\adderrad,\boxheight/2);
    \draw[very thick] (\boxwidth+\arrowlen+\arrowlen+\boxwidth+\arrowlen,\boxheight/2) node[font=\LARGE]{$+$} circle (\adderrad);
    \draw[very thick,->] (\boxwidth+\arrowlen+\arrowlen+\boxwidth+\arrowlen,\boxheight/2-\arrowlen) node[anchor=north]{$n_\mathrm{AWGN}\left( t \right)$} -- (\boxwidth+\arrowlen+\arrowlen+\boxwidth+\arrowlen,\boxheight/2-\adderrad);
    \draw[very thick, ->] (\boxwidth+\arrowlen+\arrowlen+\boxwidth+\arrowlen+\adderrad,\boxheight/2) -- (\boxwidth+\arrowlen+\arrowlen+\boxwidth+\arrowlen+\arrowlen,\boxheight/2);
    \draw[very thick] (\boxwidth+\arrowlen+\arrowlen+\boxwidth+\arrowlen+\arrowlen+\adderrad,\boxheight/2) node[font=\LARGE]{$\times$} circle (\adderrad);
    \draw[very thick, ->] (\boxwidth+\arrowlen+\arrowlen+\boxwidth+\arrowlen+\arrowlen+\adderrad,\boxheight/2-\arrowlen) node[anchor=north]{$\mathrm{e}^{\j\phi_\mathrm{RX}\left(t\right)}$\strut} -- (\boxwidth+\arrowlen+\arrowlen+\boxwidth+\arrowlen+\arrowlen+\adderrad,\boxheight/2-\adderrad);
    \draw[very thick,->] (\boxwidth+\arrowlen+\arrowlen+\boxwidth+\arrowlen+\arrowlen+\adderrad+\adderrad,\boxheight/2) -- (\boxwidth+\arrowlen+\arrowlen+\boxwidth+\arrowlen+\arrowlen+\adderrad+\arrowlen,\boxheight/2);
    \draw[ultra thick, rounded corners=\roundingrad,color=uni_apfelgruen] (\boxwidth+\arrowlen+\arrowlen+\boxwidth+\arrowlen+\arrowlen+\adderrad+\arrowlen-\boundingbox,-\boundingbox) rectangle (\boxwidth+\arrowlen+\arrowlen+\boxwidth+\arrowlen+\arrowlen+\adderrad+\arrowlen+\boxwidth+\boundingbox,\boundingbox+\boxheight);
    \draw (\boxwidth+\arrowlen+\arrowlen+\boxwidth+\arrowlen+\arrowlen+\adderrad+\arrowlen+\boxwidth/2,\boxheight+\boundingbox/2) node[anchor=south]{\bfseries\strut RX};
    \draw (\boxwidth+\arrowlen+\arrowlen+\boxwidth+\arrowlen+\arrowlen+\adderrad+\arrowlen+\boxwidth+\boundingbox,\boxheight/2) node[anchor=west]{$y\left( t \right)$};

    \draw[very thick] (\boxwidth+\arrowlen+\arrowlen+\boxwidth+\arrowlen+\arrowlen+\adderrad+\arrowlen,0) rectangle (\boxwidth+\arrowlen+\arrowlen+\boxwidth+\arrowlen+\arrowlen+\adderrad+\arrowlen+\boxwidth,\boxheight);
    \draw (\boxwidth+\arrowlen+\arrowlen+\boxwidth+\arrowlen+\arrowlen+\adderrad+\arrowlen+\boxwidth/2,\boxheight/2) node[text width=80, align=center]{Matched filter\strut\\\footnotesize $H_\mathrm{CD}^{-1}\left( f \right)$, RRC filter};

\end{tikzpicture}
  \caption{Block diagram of the communication link with both transmitter and receiver phase noise.}
  \label{fig::block_diagram}
\end{figure*}

To model the effects of EEPN, we assume a simple optical communication link.
It considers a dispersive channel including complex additive white Gaussian noise (AWGN) as the only impairments apart from the LO phase noise processes.
The block diagram of the communication link in baseband representation is shown in Fig.~\ref{fig::block_diagram}.
It is a simplified model of an optical coherent transceiver with ideal optical front-ends.
We assume that the polarization effects can be perfectly compensated and LO phase noise are independently impaired on each polarization.
This allows us to focus on a single polarization to study the penalty caused by the LO phase noise.
The transmitter (TX) will output randomly selected symbols from a 16-QAM constellation that are upsampled to the simulation frequency $\simfreq$ and pulse shaped using a root-raised cosine (RRC) filter with a roll-off of $0.1$.
This signal is then mixed with the phase noise of the transmit laser $\operatorname{exp}\left\{\im\phi_\mathrm{TX}\left( t \right)\right\}$ before passing it through the dispersive channel with transfer function $H_\mathrm{CD}\left( f \right)$.
The dispersed signal is additionally impaired by additive white Gaussian noise (AWGN), resulting in an SNR of 13.7dB.
The motivation for this specific SNR stems from the oFEC standard since the target SNR is given as $12.7\,\mathrm{dB}$.
It gives $1\,\mathrm{dB}$ room for additional impairments like EEPN.
EEPN causes some additional SNR penalty, which can be partially mitigated in DSP.
Afterwards, the signal is mixed with the phase noise of the receiver laser $\operatorname{exp}\left\{\im\phi_\mathrm{RX}\left( t \right)\right\}$.
The receiver (RX) performs matched filtering, i.e., compensating for chromatic dispersion and low-pass filtering before the signal is downsampled to the symbol rate $\symbolrate$.
In this work, perfect chromatic dispersion compensation in frequency domain is assumed  to focus on %
the impact of the phase noise.

To simplify the analytical derivation, we assume an ideal Nyquist pulse shaping even though the simulation are done using a practical RRC pulse-shaping. Assuming an ideal pulse shaping, the channel model is similar to \cite{Kakkar2015} given as
\begin{align}
  \begin{split}
    y\left( t \right) = &\mathcal{F}^{-1}\Big\{ \mathcal{F}\Big\{ \mathcal{F}^{-1}\Big\{\mathcal{F}\Big\{\\
    & x\left( t \right)\e^{\im\phi_\mathrm{TX}\left( t \right)} \Big\}H_\mathrm{CD}\left( f \right)  \Big\} \e^{\im\phi_\mathrm{RX}\left( t \right)}\Big\}H^{-1}_\mathrm{CD}\left( f \right) \Big\},
  \end{split}
  \label{eq::system_eq_time}
\end{align}
where $\mathcal{F}$ and $\mathcal{F}^{-1}$ denote the Fourier transform and its inverse, respectively.
In this, and all the following equations, we slightly abuse the notation of Fourier transform for simplicity.
The Fourier transform is of course a discrete one (DFT) as we are in discrete time.
We denote the frequencies as $f\in(-\infty,\infty)$.
However, this does not change the conclusions and the results.
The chromatic dispersion filter $H_\mathrm{CD}\left( f \right)$ is described by
\begin{align}
  H_\mathrm{CD}\left( f \right) = \e^{-\im\frac{\beta_2}{2}\left( 2\pi \right)^2f^2\ell}
  \label{eq::dispersion_filter}
\end{align}
with the group velocity dispersion parameter $\beta_2$ and the transmission length $\ell$.

\subsection{Phase Noise Model \label{sec::phase_noise_model}}
The phase noise of a laser is commonly modeled as a Wiener process \cite{Henry1986} %
given by
\begin{align}
  \phi_k= \phi_0 +  \sum\limits_{n=1}^{k} \Delta\phi_n,\;\Delta\phi_n \sim \mathcal{N}\left(0,2\pi\nicefrac{\lolw}{\simfreq}\right)
  \label{eq::wiener_process}
\end{align}
in discrete time, where the initial random starting phase of the laser is denoted by $\phi_0$ and $\lolw$ is the linewidth of the laser.
The samples $\Delta\phi_n$ are zero-mean and independent identically distributed (i.i.d.) Gaussian random variables.
We choose the samples to be on simulation frequency $\simfreq$ which can be considered as the sampling rate of the analog-to-digital converter (ADC).

A realization of such a stochastic process at any given point $k$ is zero-mean and the variance is given by
\begin{align*}
  \mathrm{Var}\left[\phi_k\right] = 2\pi k\nicefrac{\lolw}{\simfreq}.
\end{align*}
Due to the cumulative nature of the process the variance increases with $k$ and there is a correlation between the phase noise at different points in time.
The autocorrelation function of the phase noise is given by
\begin{align*}
  R_{\phi\phi}\left( k,l \right) =  2\pi \nicefrac{\lolw}{\simfreq} \min \left\{ k,l \right\}.
\end{align*}
This results in the power spectral density (PSD)
\begin{align*}
  \operatorname{PSD}_{\Phi}\left( f \right) = \frac{1}{\pi} \frac{\nicefrac{\lolw}{2}}{\left( \nicefrac{\lolw}{2} \right)^2 + f^2}.
\end{align*}
It is a Lorentzian spectrum of the phase noise, i.e., the PSD of the phase noise drops with $\nicefrac{1}{f^2}$.
The full-width half-maximum (FWHM) of the Lorentzian spectrum is defined as the linewidth $\Delta \nu$ of the laser.
It is commonly used as an indicator for the quality of the laser and its behavior as in Eq.~(\ref{eq::wiener_process}).

However, it is important to note that this model is only an approximation of the true behavior of the phase noise.
For instance, a discrepancy between the predicted performance from the simulated model and the actual performance in real-world setup is reported in \cite{Xu2023,Qiu2024}.
Recently,  it was proposed to fit the PSD by applying filters to match specific lasers \cite{Xu2024}.

Nevertheless, the Wiener process is a good starting point to understand the behavior of the phase noise.
It gives a good understanding of the predominant noise source in the laser and can be adapted as in the aforementioned paper to better match the real-world.

\section{Proposed Phase Noise Model\label{sec::proposed_phase_noise_model}}
The stochastic nature of the phase noise process as in Eq.~(\ref{eq::wiener_process}) makes it difficult to directly describe EEPN. 
Therefore, the goal of the following section is to motivate and suggest an approximated phase noise model that results in an easier description of EEPN.

The fundamental assumption underlying our derivations is that the LO phase noise process within a symmetric window of $2N+1$ samples around any time $k$ is well approximated by a linear regression function.
In other words, at time $k$, the neighboring $N$ samples on $\simfreq$ to the left and right are used for a linear regression.
These $N$ samples correspond to
\begin{align}
    N_\mathrm{S} = N\cdot\nicefrac{R_\mathrm{S}}{\simfreq}
\end{align}
symbols.
As the linewidths of the local oscillators are much smaller than the bandwidth of the symbols, the approximation is reasonable.
In the context of chromatic dispersion, the number of symbols over which the pulse is spread, for our purposes referred to as $2N_\text{CD} + 1$, can be calculated from
\begin{align}
    2N_\mathrm{CD} + 1= \left(2\left\lfloor-\pi\beta_2\ell R^2_\mathrm{S}\right\rfloor+1\right)\cdot\nicefrac{f_\mathrm{sim}}{R_\mathrm{S}}
\end{align}
The required condition is that $N_\mathrm{S}\geq N_\mathrm{CD}$ to capture the full extend of EEPN.
We verify this condition in the next subsection.

For each symbol at time $k$, we have
\begin{align}
    \phox_k\left(\lint\right) =&\; a_{1,k}\lint + a_{0,k}
\end{align}
with
$\lint\in [-N,N]$, i.e., $\lint$ is the time variable of the linear regression for the $k$th sample.
 
In this model, the subscript $k$ denotes the time-dependent nature of the affine function $\phox$ which has two parameters, namely, the slope $a_{1,k}$ and the interception $a_{0,k}$.
They are determined using a linear regression, i.e.,
\begin{align}
    a_{1,k} =& \; \frac{\sum\limits_{i=-N}^{N}i(\phi_{k+i} - \frac{1}{2N+1}\sum\limits_{l=-N}^{N}\phi_{k+l})}{\sum\limits_{i=-N}^{N}i^2}\\
    a_{0,k} =& \frac{1}{2N+1}\sum\limits_{i=-N}^{N}\phi_{k+i}.
    \label{eq::linearize_phase}
\end{align}

The slope $a_{1,k}$ can with regard to \cite{Kakkar2017} and \cite{Abolfathimomtaz2024} be seen as the mean frequency within this window and the main frequency component of the Fourier series, respectively.

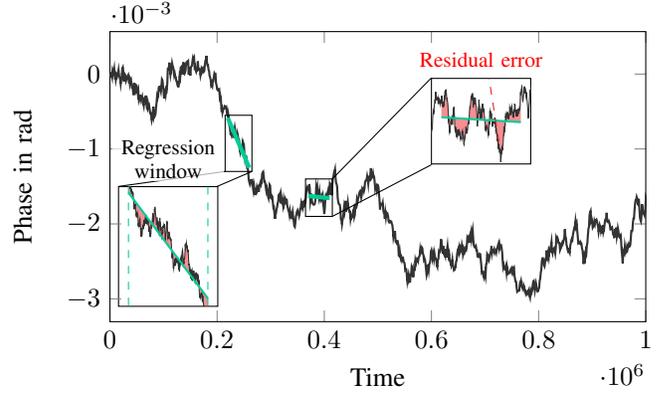
\begin{figure}
  \centering
  \begin{tikzpicture}
  \def\scalef{0.6}
  \begin{axis}[
      xmin = 0,
      xmax = 1e6,
      width=.8\textwidth*\scalef,
      height=.5\textwidth*\scalef,
      xlabel={Time},
      xlabel style={yshift=1},
      ylabel={Phase in rad},
      ylabel style={yshift=-2},
    ]
    \addplot+[thick,each nth point=100, mark=none,unigrau] table[col sep=comma] {fig/linearize_phase/wiener_process.csv};
    \addplot[ultra thick,plotlygreen] table[col sep=comma] {fig/linearize_phase/slope_240000.csv};
    \addplot[ultra thick,plotlygreen] table[col sep=comma] {fig/linearize_phase/slope_390000.csv};

    \coordinate (plot1) at (axis cs: 16000,-3.1e-3); %
    \coordinate (plot2) at (axis cs:600000,-1.2e-3); %

    \draw (axis cs: 215000,-0.55e-3) rectangle (axis cs: 265000,-1.3e-3);
    \draw (axis cs: 365000, -1.9e-3) rectangle (axis cs: 415000, -1.4e-3);

    \draw (axis cs: 215000,-1.3e-3) -- (axis cs: 16000, -1.5e-3);
    \draw (axis cs: 265000,-1.3e-3) -- (axis cs: 200000, -1.5e-3);

    \draw (axis cs: 415000,-1.4e-3) -- (axis cs: 600000, -0.05e-3);
    \draw (axis cs: 415000,-1.9e-3) -- (axis cs: 600000, -1.21e-3);
  \end{axis}
  \begin{axis}[
      at={(plot1)},
      xmin = 215000,
      xmax = 265000,
      ymax=-0.55e-3,
      ymin=-1.3e-3,
      width=.8\textwidth*0.2,
      height=.5\textwidth*0.35,
      scaled x ticks = false,
      scaled y ticks = false,
      xtick = {0},
      xticklabels = {,,},
      ytick = {0},
      yticklabels = {,,},
      title={Regression window},
      title style={font=\footnotesize,yshift=-4, fill=white, fill opacity=.7,text opacity=1, text width=40,align=center,inner sep=1}
    ]
    \addplot+[each nth point=10, mark=none,unigrau,name path=slopewinaa] table[col sep=comma] {fig/linearize_phase/wiener_process_slope_240000.csv};
    \addplot+[thick,plotlygreen, mark=none, name path=slopeaa] table[col sep=comma] {fig/linearize_phase/slope_240000.csv};

    \addplot[fill=uni_rot, fill opacity=0.5] fill between[of=slopewinaa and slopeaa, soft clip={domain=220000:260000}];

    \draw[dashed, plotlygreen] (axis cs: 220000,-0.55e-3) -- (axis cs: 220000,-1.3e-3);
    \draw[dashed, plotlygreen] (axis cs: 260000,-0.55e-3) -- (axis cs: 260000,-1.3e-3);
  \end{axis}
  \begin{axis}[
      at={(plot2)},
      xmin = 365000,
      xmax = 415000,
      width=.8\textwidth*0.2,
      height=.5\textwidth*0.3,
      scaled x ticks = false,
      scaled y ticks = false,
      xtick = {0},
      xticklabels = {,,},
      ytick = {0},
      yticklabels = {,,},
      ymin=-1.9e-3,
      ymax=-1.4e-3,
      title={Residual error},
      title style={color=red, font=\footnotesize,yshift=-3, fill=white, fill opacity=.7, text opacity=1, inner sep=1}
    ]
    \addplot+[each nth point=10, mark=none,unigrau, name path=slopewinb] table[col sep=comma] {fig/linearize_phase/wiener_process_slope_390000.csv};
    \addplot+[thick,plotlygreen, mark=none, name path=slopeb] table[col sep=comma] {fig/linearize_phase/slope_390000.csv};
    \addplot[fill=uni_rot, fill opacity=0.5] fill between[of=slopewinb and slopeb, soft clip={domain=370000:410000}];

    \draw[dashed, uni_rot] (axis cs: 399000, -1.77e-3) -- (axis cs: 394000, -1.4e-3);
  \end{axis}
\end{tikzpicture}
  \vspace*{-.3cm}
  \caption{Illustration of the sliding-window based linear regression for an exemplary random walk. The regression window size is 5200 symbols. The red shaded areas indicate the residual error of the approximation.}
  \label{fig::linearize_phase}
\end{figure}

A visual representation of this idea is depicted in Fig.~\ref{fig::linearize_phase}.
There, an exemplary Wiener process is linearized around two different points in time.
For the linearization, the windowed regression as in Eq.~(\ref{eq::linearize_phase}) is applied for each time step.
Due to the windowed nature which takes neighboring samples into account, the slopes at different points in time are correlated.

With the linearization, we can describe the phase noise around time $k$ as
\begin{align}
    \e^{\im \phi_k}\Bigr\vert_{k\in\left[-N,N\right]} \approx \e^{\im \phox_k\left(\lint\right)} = \e^{\im \left(a_{1,k}\lint + a_{0,k}\right)}.
\end{align}

This function can be transformed into the frequency domain with respect to $\lint$ using the shifting property of the Fourier transform
\begin{align}
    \mathcal{F}_{\lint}\left\{  \e^{\im \phox_k\left(\lint\right)}\right\} = \delta\left(f - \nicefrac{a_{1,k}}{2\pi}\right)\e^{\im a_{0,k}}.
\end{align}
Inserting this result into the system equation Eq.~(\ref{eq::system_eq_time}) for the LO phase noise at TX and RX results in
\begin{align}
    \begin{split}
        \tilde{y}_k\left(\lint\right) = \e^{\im (\txinterception + \rxinterception)}\mathcal{F}_{\lint}^{-1}\Bigg\{  &X_k\left(f - \nicefrac{\txslope}{2\pi} - \nicefrac{\rxslope}{2\pi}\right) \\
        &H_\mathrm{CD}\left(f - \nicefrac{\rxslope}{2\pi}\right)  H_\mathrm{CD}^{-1}\left(f\right)  \Bigg\},
    \end{split}
\end{align}
where $\txslope$ and $\txinterception$ and $\rxslope$ and $\rxinterception$ refer to the regression parameters of TX and RX phase noise, respectively.
The subscript $k$ in $X_k$ is used to indicate that the Fourier transform is valid only within the window.
For $\lint$ it holds that
\begin{align}
    \tilde{y}_k\left(0\right) \approx y_k.
\end{align}
With this formulation two aspects arising from the employed model are noticeable.

Primarily, the discrepancy between the dispersion compensation and the channel becomes evident.
This discrepancy stems from the frequency shift as modeled by the linearization, causing the parabolic phase of the dispersive channel as in Eq.~(\ref{eq::dispersion_filter}) and its compensation filter to become misaligned.
Consequently, this misalignment leads to the presence of artifacts in the received signal.
This CDC discrepancy can be, similar to \cite{Kakkar2015}, rewritten with Eq.~(\ref{eq::dispersion_filter}) to the form
\begin{align}
    H_\mathrm{CD}(f-\nicefrac{\rxslope}{2\pi})H_\mathrm{CD}^{-1}\left(f\right) =\e^{-\im\frac{\beta_2}{2} \rxslopesq \ell} \e^{-\im\beta_2 2\pi f\rxslope\ell}.
\end{align}
Applying the inverse Fourier transform yields
\begin{align}
    \mathcal{F}^{-1}\Big\{ H_\mathrm{CD}(f-\nicefrac{\rxslope}{2\pi})H_\mathrm{CD}^{-1}\left(f\right) \Big\} = \e^{-\im\frac{\beta_2}{2} \rxslopesq \ell}\delta_{\lint + \beta_2 \rxslope \ell}
\end{align}

The received signal with these assumptions is given by
\begin{align}
    \begin{split}
    y_k\left(\lint\right) =& \e^{\im(\txslope \lint + \txinterception)}\e^{\im(\rxslope \lint  + \rxinterception)}\\
    &x_{k + \beta_2 \rxslope \ell}\e^{\im\beta_2\rxslope\txslope\ell}\e^{\im\frac{\beta_2}{2}\rxslopesq \ell}
    \end{split}
\end{align}
Evaluating only the values that are relevant for the current time $k$, i.e., $\lint = 0$ gives 
\begin{align}
    y_k = \e^{\im (\txinterception + \rxinterception)} x_{k + \beta_2 \rxslope \ell}\e^{\im\beta_2\rxslope\txslope\ell}\e^{\im\frac{\beta_2}{2}\rxslopesq \ell}
    \label{eq::received_signal_linear}
\end{align}
which suggests that EEPN induces a timing error that is related to the (long term) slope of the RX LO phase noise process.
We will use this observation to verify our assumptions in the following subsection.

\subsection{Verification of the assumptions \label{sec::linearization_verification}}
We performed simulations with the goal to obtain the
timing error and relate it to the slope of the RX LO phase noise process.
Due to the stochastic nature of phase noise process, there is no deterministic way to estimate the timing error.
To avoid algorithmic effects and penalties, a genie-aided timing recovery algorithm was used, giving us the exact sampling position.
For this the transmitted and received symbols were upsampled by a factor of $200$ and correlated in a windowed manner.
This correlation then allowed to detect the shift that gives the maximum agreement between sent and received symbols giving us the ``ground-truth'' timing error.
Simulations were performed for an AWGN-free channel.
Since the calculations of the timing error are performed genie-aided, this does not change the applicability to noisy channels.

\begin{figure}
  \centering
  \begin{tikzpicture}
  \def\scalef{0.6}
  \def\displen{1362-100}
  \begin{groupplot}[group style={group size= 1 by 3, vertical sep=.3cm}]
    \nextgroupplot[
      xmin=50, xmax=6050,
      ymin=-0.05, ymax=1.05,
      xtick={0, 1000, 2000, 3000, 4000, 5000, 6000},
      ytick={0, 0.2, 0.4, 0.6, 0.8, 1},
      width=.8*\textwidth*\scalef, height=.5*\textwidth*\scalef,
      xticklabels={,,},
      ylabel={Pearson coefficient},
      tick align=outside,
      tick pos=left,
      ylabel style = {yshift=-.2cm},
    ]
    \addplot graphics [includegraphics cmd=\pgfimage,xmin=50, xmax=6050, ymin=-0.05, ymax=1.05] {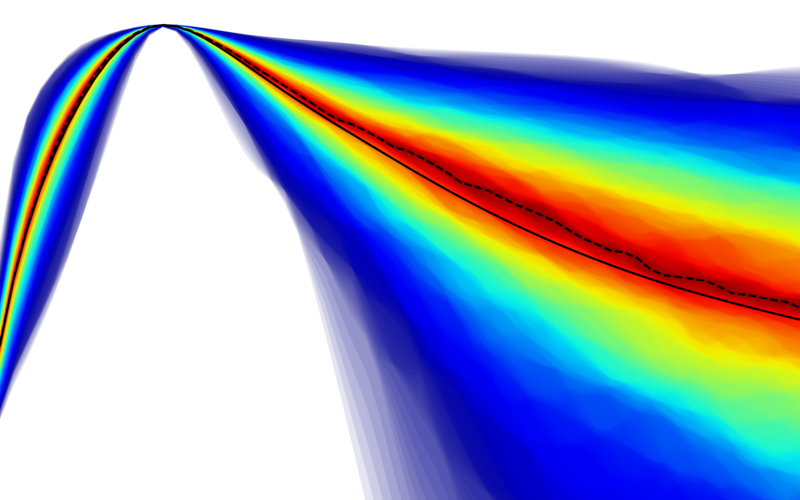};
    \draw[line width=2,unigrau,opacity=.7] (axis cs: \displen, -.05) -- (axis cs: \displen, 1.05);
    \draw (axis cs: \displen, .5) node[rotate=90, yshift=4,font=\footnotesize]{CD memory\strut};
    \coordinate (legend_2000) at (axis description cs:0.995,.015); %
    \nextgroupplot[
      xmin=50, xmax=6050,
      ymin=-0.05, ymax=1.05,
      xtick={0, 1000, 2000, 3000, 4000, 5000, 6000},
      ytick={0, 0.2, 0.4, 0.6, 0.8, 1},
      width=.8*\textwidth*\scalef, height=.5*\textwidth*\scalef,
      xticklabels={,,},
      ylabel={Pearson coefficient},
      tick align=outside,
      tick pos=left,
      ylabel style = {yshift=-.2cm},
    ]
    \addplot graphics [includegraphics cmd=\pgfimage,xmin=50, xmax=6050, ymin=-0.05, ymax=1.05] {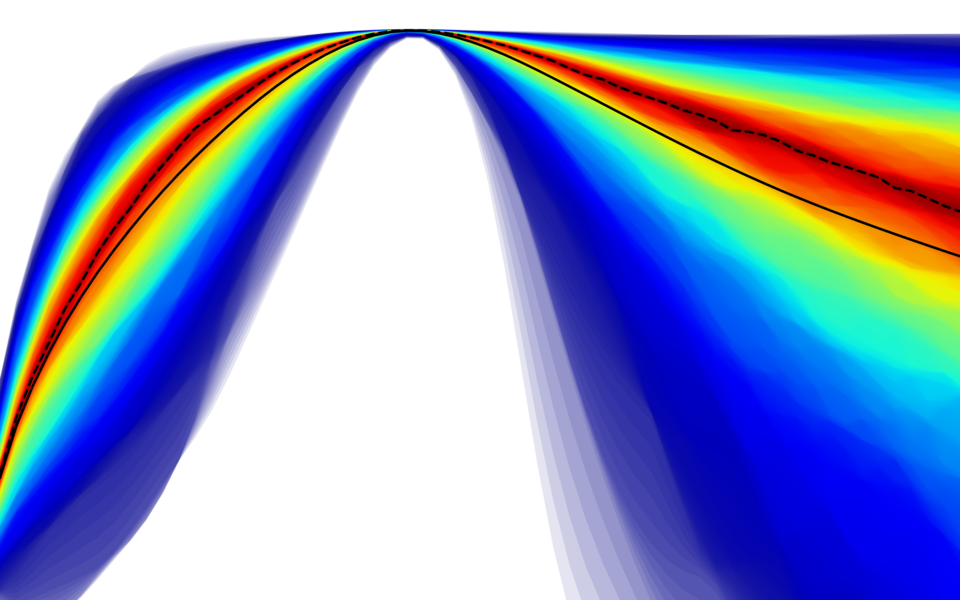};
    \draw[line width=2,unigrau,opacity=.7] (axis cs: 2723-100, -.05) -- (axis cs: 2723-100, 1.05);
    \draw (axis cs: 2723-100, .5) node[rotate=90, yshift=4,font=\footnotesize]{CD memory\strut};
    \coordinate (legend_4000) at (axis description cs:0.995,.015); %
    \nextgroupplot[
      xmin=50, xmax=6050,
      ymin=-0.05, ymax=1.05,
      xtick={0, 1000, 2000, 3000, 4000, 5000, 6000},
      ytick={0, 0.2, 0.4, 0.6, 0.8, 1},
      width=.8*\textwidth*\scalef, height=.5*\textwidth*\scalef,
      xlabel={Half window length $N_\mathrm{S}$},
      ylabel={Pearson coefficient},
      tick align=outside,
      tick pos=left,
      ylabel style = {yshift=-.2cm},
    ]
    \addplot graphics [includegraphics cmd=\pgfimage,xmin=50, xmax=6050, ymin=-0.05, ymax=1.05] {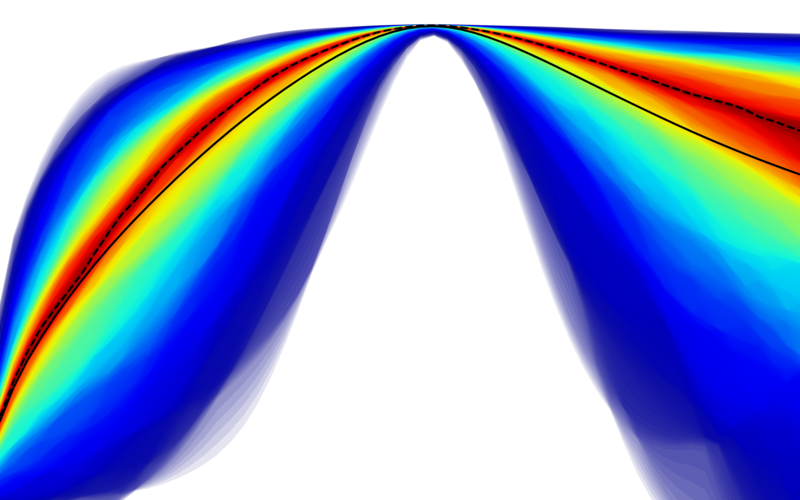};
    \draw[line width=2,unigrau,opacity=.7] (axis cs: 3404-100, -.05) -- (axis cs: 3404-100, 1.05);
    \draw (axis cs: 3404-100, .5) node[rotate=90, yshift=4,font=\footnotesize]{CD memory\strut};
\coordinate (legend_5000) at (axis description cs:0.995,.015); %
  \end{groupplot}

  \matrix [
    draw,
    fill=white,
    fill opacity=.8,
    text opacity=1,
    matrix of nodes,
    align =left,
    row sep = -1,
    column sep = -8,
    inner sep= 3,
    text width=40,
    anchor=south east,
    font=\footnotesize,
    mark options={solid},
    rounded corners=4pt
  ] (mymat) at (legend_2000) {
  Length:&$2000\,\mathrm{km}$\\
  Linewidth:&$500\,\mathrm{kHz}$\\
  Dispersion:&$34\,\nicefrac{\mathrm{ns}}{\mathrm{nm}}$\\
  };
  \matrix [
    draw,
    fill=white,
    fill opacity=.8,
    text opacity=1,
    matrix of nodes,
    align =left,
    row sep = -2,
    column sep = -8,
    inner sep= 3,
    text width=40,
    anchor=south east,
    font=\footnotesize,
    mark options={solid},
    rounded corners=4pt
  ] (mymat) at (legend_4000) {
  Length:&$4000\,\mathrm{km}$\\
  Linewidth:&$300\,\mathrm{kHz}$\\
  Dispersion:&$68\,\nicefrac{\mathrm{ns}}{\mathrm{nm}}$\\
  };
  \matrix [
    draw,
    fill=white,
    fill opacity=.8,
    text opacity=1,
    matrix of nodes,
    align =left,
    row sep = -2,
    column sep = -8,
    inner sep= 3,
    text width=40,
    anchor=south east,
    font=\footnotesize,
    mark options={solid},
    rounded corners=4pt
  ] (mymat) at (legend_5000) {
  Length:&$5000\,\mathrm{km}$\\
  Linewidth:&$150\,\mathrm{kHz}$\\
  Dispersion:&$85\,\nicefrac{\mathrm{ns}}{\mathrm{nm}}$\\
  };
\end{tikzpicture}
  \caption{Pearson coefficient of genie-aided timing error estimate and slope from linear regression over different window sizes $N_\mathrm{S}$ in symbols at rate $R_\mathrm{S}=100\,\mathrm{GBd}$ showing very good agreement between the two. The CD memory (in taps of duration $\nicefrac{1}{R_\mathrm{S}}$) is given as a reference to the respective fiber length.}
  \label{fig::correlate_slope_genie_aided}
\end{figure}
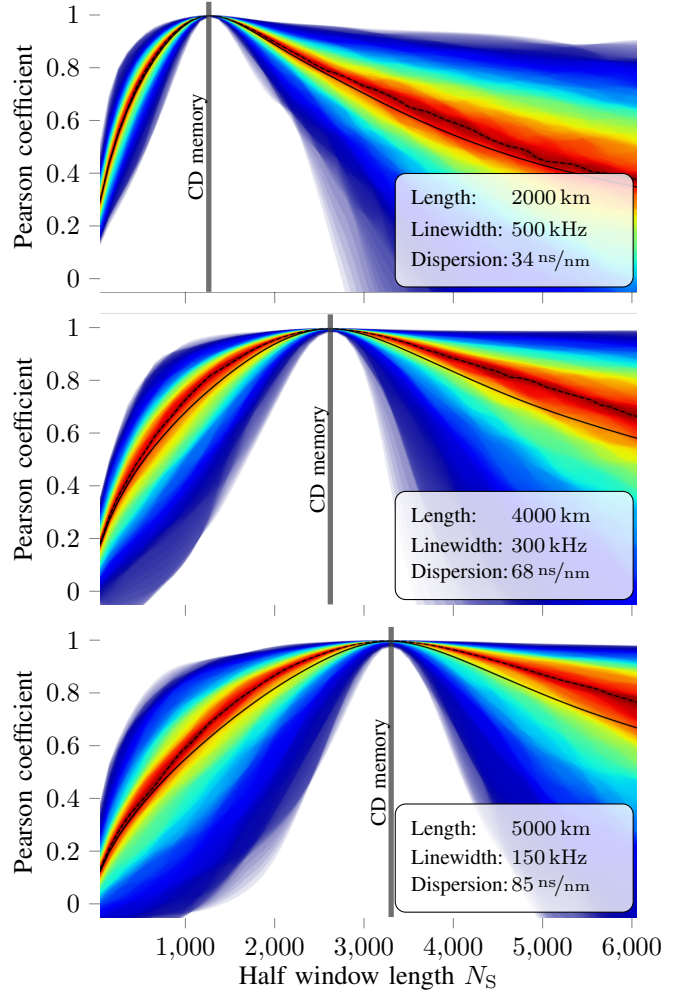

The ``ground-truth'' timing error estimated this way was then, in turn, correlated with the slope of the receiver phase noise process.
The slope was estimated in a genie-aided manner as well, i.e., the phase noise samples were known and a windowed linear regression over different window sizes $N$ was performed.
Both timing error estimations were then correlated and the Pearson coefficient was calculated.
For each window size, $200$ simulations with $25000$ symbols each were performed.
The results can be seen in Fig.~\ref{fig::correlate_slope_genie_aided} for three different linewidths - length combinations: $500\,\mathrm{kHz}$ and $2000\,\mathrm{km}$, $300\,\mathrm{kHz}$ and $4000\,\mathrm{km}$, and $150\,\mathrm{kHz}$ and $5000\,\mathrm{km}$.
This selection ensures that we have different amounts of accumulated dispersion while keeping the EEPN penalty to some degree similar.
Hence, the plots are comparable.
On the $y$-axis, the Pearson coefficient and on the $x$-axis the half window size $N_\mathrm{S}$ in symbols is shown.
Here, $\nicefrac{f_\mathrm{sim}}{R_\mathrm{S}}\left(2N_\mathrm{S}+1\right)$ phase noise samples were used in the simulation.
In the plots, the relative density of the Pearson coefficients is depicted.

In Fig.~\ref{fig::correlate_slope_genie_aided}, we can see that the timing error is very correlated to the slope.
The distribution of correlation values depends on linewidth and $N_\mathrm{S}$.
As $N_\mathrm{S}$ approaches its optimal value, the correlation concentrates and approaches 1. 
It means that around optimal $N_\mathrm{S}$, the correlation is always very close to 1 and therefore, the timing error is mainly a function of the slopes.
However, there still seems to be dependence on the linewidth in terms of the spread of the density before the optimum value.
With increasing linewidth, the range of possible values increases.

Another interesting aspect is the asymmetry of the curve.
It rises steeply with increasing window size.
After reaching the optimum, however, the range of possible Pearson coefficients spreads through the whole range of possible values ss samples outside of the scope of the chromatic dispersion memory are taken into account.
However, these samples are irrelevant for the effect of EEPN as it stems from the interaction of the LO phase noise with EDC.

This gives a good intuition for selecting the linear regression window and validating the assumption of the proposed model.
The results show that, if $N_\mathrm{S}$ is selected carefully, the timing error can be characterized mainly as a function of the ``instantaneous'' slope of the phase noise.
The mean frequency of the phase noise process is the most significant frequency component for EEPN induced penalties before timing recovery.
Of course, the current model makes errors due to the linearization.
In the next section, this residual error will be evaluated and described.
It allows us to split EEPN into four separate terms, which is advantageous for further analysis.

\section{Residual Error\label{sec::residual_error}}
In Fig.~\ref{fig::linearize_phase}, the error incurred by the linearization process as proposed in Eq.~(\ref{eq::linearize_phase}) is visualized by the filled area between the affine approximation and the exemplary random process.
It can be seen that - while the dominant effect is well captured - it remains yet a non-negligible second-order error.
Therefore, we now want to include this residual error into our model and find a (stochastic) description for it.
In the most basic form, we can write
\begin{align*}
    \e^{\im\phi_k} = \e^{\im\phox_k\left(\lint\right) + \im n_k\left(\lint\right)},
\end{align*}
where $n_k\left(\lint\right)$ is the residual error given by
\begin{align*}
    n_k\left(\lint\right) = \phi_{k+\lint} - a_{0,k} - a_{1,k}\lint
\end{align*}
with the regression time $\lint$.

Assuming that $n_k\left(\lint\right)$ is just a small perturbation, we can approximate the actual $\phi_k$ using the first-order Taylor expansion as follows
\begin{align}
    \e^{\im\phi_{k}\left(\lint\right)} \approx \e^{\im\phox_k\left(\lint\right)}\left(1 + \im n_k\left(\lint\right)\right).
\end{align}
With this simplified formulation we can again employ the shifting properties of the Fourier transform as before.
After Fourier transform with respect to $\lint$, this gives rise to
\begin{align}
    \mathcal{F}_\lint\left\{\e^{\im\phi_k\left(\lint\right)}\right\} \approx \e^{\im a_{0,k}}\left(\delta\left(f - \frac{a_{1,k}}{2\pi}\right) + \im N_k\left(f-\frac{a_{1,k}}{2\pi}\right)\right).
    \label{eq::linearize_phase_taylor}
\end{align}
From this, we can see that the our previous derivation is preserved by the first term and the respective result in Eq.~(\ref{eq::received_signal_linear}) still part of the more precise EEPN description.
We now have a new term $N_k$ which is a frequency shifted noise in the current window.

We will show a description of $N_k$ later, but first, we want to derive the formulation of the received signal with this new approximation.
The steps are similar to before and therefore not presented in full detail here.
In frequency domain, the received signal $\tilde{y}_k$ for a given $k$ within a window of $N$ is approximated by
\begin{align}
  \begin{split}
    \tilde{Y}_k&\left( f \right) = \e^{\im \left(\txinterception + \rxinterception\right)} \Big[\\
      \rotatebox{90}{\hspace*{-.5em}\Xte}&\;\Bigl\{ X_k\left( f - \frac{\txslope + \rxslope }{2\pi} \right)H_\mathrm{CD}\left( f - \frac{\rxslope }{2\pi} \right)H_\mathrm{CD}^{-1}\left( f \right) \\
      \rotatebox{90}{\hspace*{-.5em}\Nro}&\begin{cases}
        +  \im \left[X\left( f - \frac{\rxslope }{2\pi} \right) * N_{\mathrm{TX},k}\left( f - \frac{\txslope }{2\pi} - \frac{\rxslope }{2\pi} \right)\right] \\
        \;\;\;\cdot H_\mathrm{CD}\left( f - \frac{\rxslope }{2\pi} \right) H_\mathrm{CD}^{-1}\left( f \right)                                            \\
      \end{cases}\\
      \rotatebox{90}{\hspace*{-.5em}\Nrr}&\begin{cases}
        +  \im \left(\left[X\left( f - \frac{\txslope }{2\pi} \right)H_\mathrm{CD}\left( f \right)\right]*N_{\mathrm{RX},k}\left( f - \frac{\rxslope }{2\pi} \right)\right) \\
        \;\;\;\cdot H_\mathrm{CD}^{-1}\left( f \right)                                                                                                                  \\
      \end{cases}\\
      \rotatebox{90}{\hspace*{-.5em}\Nxr} &\begin{cases}
        -  \Big(\left[\left(X_k\left( f \right) * N_{\mathrm{TX},k}\left( f - \frac{\txslope }{2\pi}\right)\right)H_\mathrm{CD}\left( f \right)\right] \\
        \;\;\;*N_{\mathrm{RX},k}\left( f - \frac{\rxslope }{2\pi} \right)\Big)H_\mathrm{CD}^{-1}\left( f \right)                                     \\
      \end{cases}\\
      \Big],&
  \end{split}
  \label{eq::EEPN_model_frequency_domain}
\end{align}
where $N_{\mathrm{TX},k}$ and $N_{\mathrm{RX},k}$ refer to the Fourier transforms of the residual timing errors of the transmit and receive laser as in Eq.~(\ref{eq::linearize_phase_taylor}), respectively.
As before, the Fourier transform is calculated with respect to the regression time $\lint$.
Applying the inverse Fourier transform and setting $\lint=0$ to arrive at the result for time $k$ we get
\begin{align}
  \begin{split}
    \tilde{y}_k\left(0\right) &= \e^{\im \left( \txinterception + \rxinterception \right)}\Bigg[\\
      \rotatebox{90}{\hspace*{-.5em}\xte}&\Bigl\{ x_{k+\beta_2 \rxslope \ell}\cdot\e ^{\im \beta_2 \txslope \rxslope \ell}\e^{\im \frac{\beta_2}{2}\rxslope ^2\ell}\\
      \rotatebox{90}{\hspace*{-.5em}\nro}&\begin{cases}+\im x_{k + \beta_2 \rxslope \ell}\cdot n_{\mathrm{TX}, k + \beta_2 \rxslope  \ell}\\
      \;\;\;\cdot\e^{\im \beta_2 \txslope \rxslope \ell}\e^{\im \frac{3\beta_2}{2}\rxslope ^2 \ell}\end{cases}\\
      \rotatebox{90}{\hspace*{-.5em}\nrr}&\begin{cases}+\im \sum\limits_{f_1=-\infty}^{\infty}x_{k + \beta_2 2\pi f_1 \ell}\cdot \e^{\im \beta_2 \txslope 2\pi f_1 \ell}\\
      \;\;\;\;\;\;\;\;\;\;\;\cdot\e^{\im \frac{\beta_2}{2}\left( 2\pi \right)^2f_1^2\ell}N_\mathrm{RX}\left( f_1 - \frac{\rxslope }{2\pi} \right)%
      \end{cases}\\
      \rotatebox{90}{\hspace*{-.5em}\nxr}&\begin{cases}-\sum\limits_{f_1=-\infty}^{\infty}x_{k + \beta_2 2\pi f_1 \ell}\cdot\e^{\im \frac{3\beta_2}{2} \left( 2\pi \right)^2 f_1^2 \ell}\\
      \;\;\;\;\;\;\;\;\;\;\;\cdot n_{\mathrm{TX}, k + \beta_2 2\pi f_1 \ell}\cdot \e^{\im \beta_2 \txslope 2\pi f_1 \ell}\\
      \;\;\;\;\;\;\;\;\;\;\;\cdot N_\mathrm{RX}\left( f_1 - \frac{\rxslope }{2\pi} \right)\\
      \end{cases}\\
      &\Bigg].
  \end{split}
  \label{eq::EEPN_model_time_domain_zero}
\end{align}

As indicated in Eq.~(\ref{eq::EEPN_model_frequency_domain}) and (\ref{eq::EEPN_model_time_domain_zero}) this enables to describe EEPN by four separate terms, each with its own distinct properties.
We call them \emph{timing error term} $\xte \trf \Xte$, \emph{rotation term} $\nro \trf \Nro$, \emph{receiver residual noise term} $\nrr \trf \Nrr$, and \emph{cross residual noise term} $\nxr \trf \Nxr$ with the symbols in time and frequency domain separated by the transformation symbols.
Denoting the timing error term with $x$ instead of $n$ for noise captures that most information about the transmitted signal is contained within this whereas the three other terms only contain little to no information.
With this, we can approximate the received signal with EEPN penalty and write Eq.~(\ref{eq::EEPN_model_frequency_domain}) and (\ref{eq::EEPN_model_time_domain_zero}) as
\begin{align*}
\begin{split}
    Y_k\left(f\right) = \e^{\im\varphi_{0,k}}\Bigl[&{\Xte}_{,k}\left(f\right) + {\Nro}_{,k}\left(f\right)\\&+ {\Nrr}_{,k}\left(f\right) + {\Nxr}_{,k}\left(f\right)\Bigr]
\end{split}
\end{align*}
and 
\begin{align}
    y_k = \e^{\im\varphi_{0,k}}\left[{\xte}_{,K} + {\nro}_{,k} + {\nrr}_{,k} + {\nxr}_{,k}\right].
\end{align}
Here, we combined the interceptions of TX and RX LO phase noise process to $\varphi_{0,k}=\txinterception + \rxinterception$.
The different terms in time domain are visualized in Fig.~\ref{fig::EEPN_terms}, where the impact per symbol index is color coded to visualize the dependency of the respective terms.

With this, we are able to better understand EEPN and the influence of different DSP algorithms on the mitigation of it which will be discussed in Section~\ref{sec::dsp_aware_analysis}

\subsection{Verification of the model}
\begin{figure*}
  \centering
  \input{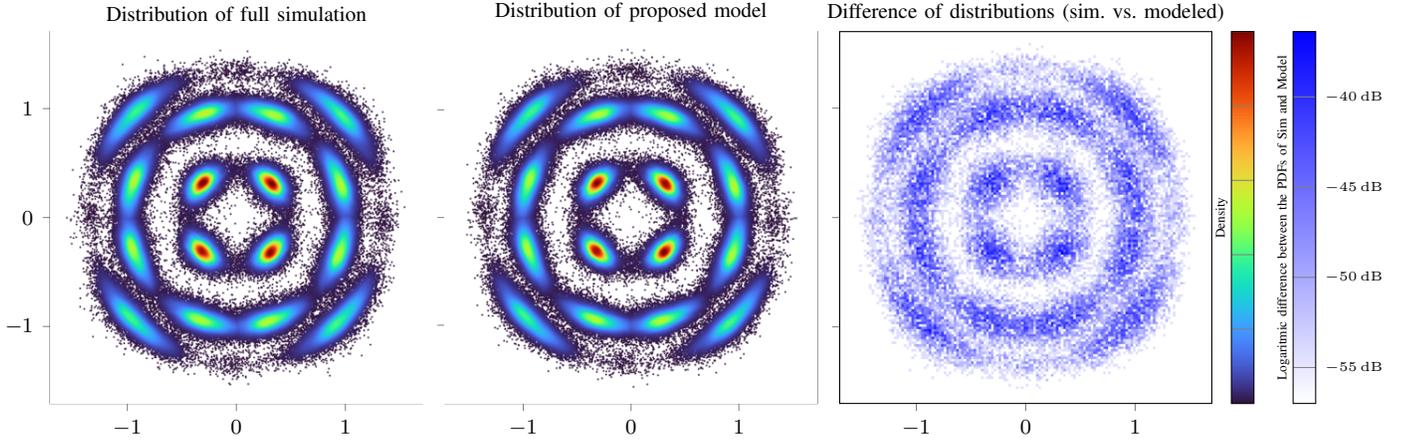}
  \caption{Comparison of the distribution of full system simulation and the proposed model for 300000 symbols at $150\,\mathrm{kHz}$ linewidth showing very good agreement.}
  \label{fig::model_vs_simulation}
\end{figure*}
In order to arrive at Eq.~(\ref{eq::EEPN_model_time_domain_zero}), we made some assumptions that need to be verified.
It was assumed that the LO phase noise process can be approximated within a window of $2N_\mathrm{S}+1$ symbols by an affine function.
This was already shown to be valid in Section~\ref{sec::linearization_verification} where the ``ground-truth'' timing error and the error as per the model shown perfect correlation for $N_\mathrm{S} = N_\mathrm{CD}$.

Next in the derivation, we performed a Taylor series expansion of $\exp\left\{\im n_k\left(\lint\right)\right\}$ around zero.
This was based on the assumption that the residual error is zero-mean and small ($\ll 1$).
The first assumption is easily verifiable.
The residual error can be written as
\begin{align*}
  \mathrm{E}\left[ n_k \right]  & = \mathrm{E}\left[ \phi_k \right] - \frac{1}{2N+1}\sum_{i=-N}^{N}\mathrm{E}\left[ \phi_{k+i} \right] = 0. %
\end{align*}
As the underlying process is zero-mean, this also holds for the derived process.
For the second one, the variance or standard deviation is needed.
The variance is found to be
\begin{align*}
  \nonumber
  \mathrm{Var}\left(n_k\right) & = \mathrm{E}\left[n_k^2\right] %
                                  = \frac{4\pi\Delta \nu}{\simfreq}\frac{\nicefrac{2}{3}N^3 + N^2+\nicefrac{1}{3}N}{\left(2N+1\right)^2}
\end{align*}
for the window of $N$.
Exemplary values for the standard deviation for $N_\mathrm{CD}$ for the accumulated dispersion as in Fig.~\ref{fig::correlate_slope_genie_aided} are 0.119, 0.131, and 0.1034 for the combinations $2000\,\mathrm{km}$@$500\,\mathrm{kHz}$,  $4000\,\mathrm{km}$@$300\,\mathrm{kHz}$, and $5000\,\mathrm{km}$@$150\,\mathrm{kHz}$.
This small standard deviation indicates that most samples of $n_k$  concentrate around zero resulting in a small residual error by the first-order Taylor expansion.
Hence, all assumptions seem to be applicable.

As a last step, we want to show the difference in distribution of the received symbols between a fully simulated communication link and our derived model with its assumptions and approximations.
This is shown in Fig.~\ref{fig::model_vs_simulation}.
The distribution of the received symbols is depicted for the simulated system and our proposed model in the two leftmost subfigures, respectively.
For the comparison, over $300\,000$ symbols were simulated which gives roughly $20\,000$ symbols per constellation point.
At first sight it is hard to make a distinction between the distribution of the simulation in the left-most plot and the model-based one in the center.
When looking more closely very subtle differences can be seen.
Most notably near the top left symbol where in the simulation the points are scattered further out and more sparse.
But overall, the two plots look very similar in terms of their distribution.
This first impression is supported by the difference between the distributions as shown in the right-most plot.
There, the rate of occurrence of a difference between the two distributions is shown on a logarithmic scale.
Overall the rate is quite low meaning that the two distributions are very similar.

In other words, the assumptions motivating the Taylor series expansion seem to be well justified and the model can be used to further investigate EEPN.
Furthermore, it needs to be stressed once again, that the Wiener process, which we now are able to describe in a different way, is in itself also only an approximation and does not necessarily reflect the behavior of a real-world laser \cite{Xu2023,Qiu2024}.
But the model provides us with valuable insights into EEPN and allows to understand the impact of different DSP algorithms as we will show in Sec. \ref{sec::dsp_aware_analysis}

\subsection{Analysis of residual error}
In this section, we want to briefly discuss the temporal and spectral properties of the residual error.
Autocorrelation (ACF) and power spectral density (PSD) give insights into those properties.

We derived the ACF for the residual error and give the result in Eq.~(\ref{eq::residual_acf}) on top of the next page.
\begin{figure*}[t!]
  \begin{align}
    \begin{split}
      R_{nn}\left( l \right)=& 2\pi\nicefrac{\lolw}{\simfreq}\max\left\{ 0, N+1-l \right\}\\
      &+ \frac{2\pi\nicefrac{\lolw}{\simfreq}}{\left( 2N+1 \right)^{2}}\begin{cases}
        \frac{\left|l\right|^3}{6} + l^2\left( N+\frac{1}{2} \right) + \left|l\right|\left( 2N^2 + 4N + \frac{4}{3} \right) - \frac{10}{3}N^3 - 7N^2 - \frac{14}{3}N - 1, & 0\leq \left|l\right| < N     \\
        \frac{\left|l\right|^3}{6} - l^2\left( N+\frac{1}{2} \right) + \left|l\right|\left( 2N^2 + 2N + \frac{2}{3} \right) - \frac{4}{3}N^3 - 2N^2 - \frac{2}{3}N ,      & N\leq \left|l\right| \leq 2N \\
        0,                                                                                                                                                                & \left|l\right| > 2N
      \end{cases}
    \end{split}
    \label{eq::residual_acf}
  \end{align}
\end{figure*}
The full derivation can be found in Appendix \ref{subsec::autocorrelation_derivation}.

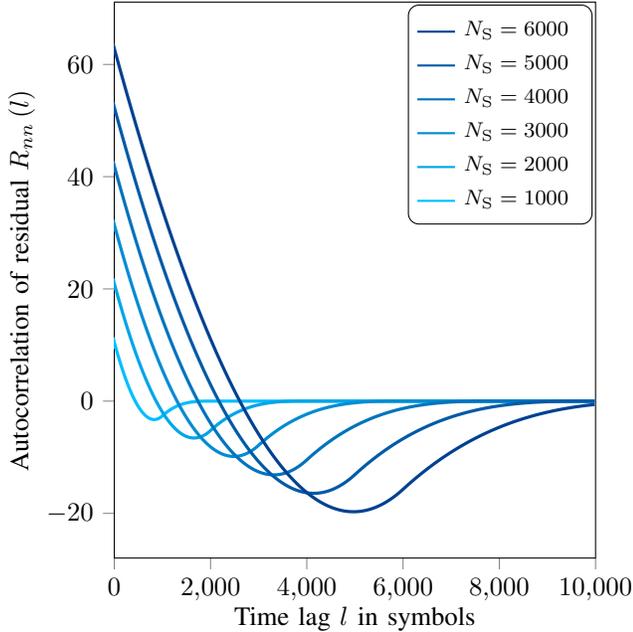
\begin{figure}%
  \centering
  \begin{tikzpicture}
  \begin{axis}[
      width=.8*\textwidth*0.55, height=.9*\textwidth*0.55,
      xlabel={Time lag $l$ in symbols},
      ylabel={Autocorrelation of residual $R_{nn}\left( l \right)$},
      tick align=outside,
      tick pos=left,
      xmin=0, xmax=10000,
      scaled x ticks=false,
    ]
    \addplot [very thick,uniblauHell!100!uniblauDunkel,mark=none, mark size=1.5, mark options={solid}, x filter/.code={\pgfmathparse{#1/1e1}}]
    table[col sep=comma] {fig/residual_acf/10000.csv};
    \addplot [very thick,uniblauHell!80!uniblauDunkel,mark=none, mark size=1.5, mark options={solid}, x filter/.code={\pgfmathparse{#1/1e1}}]
    table[col sep=comma] {fig/residual_acf/20000.csv};
    \addplot [very thick,uniblauHell!60!uniblauDunkel,mark=none, mark size=1.5, mark options={solid}, x filter/.code={\pgfmathparse{#1/1e1}}]
    table[col sep=comma] {fig/residual_acf/30000.csv};
    \addplot [very thick,uniblauHell!40!uniblauDunkel,mark=none, mark size=1.5, mark options={solid}, x filter/.code={\pgfmathparse{#1/1e1}}]
    table[col sep=comma] {fig/residual_acf/40000.csv};
    \addplot [very thick,uniblauHell!20!uniblauDunkel,mark=none, mark size=1.5, mark options={solid}, x filter/.code={\pgfmathparse{#1/1e1}}]
    table[col sep=comma] {fig/residual_acf/50000.csv};
    \addplot [very thick,uniblauHell!0!uniblauDunkel,mark=none, mark size=1.5, mark options={solid}, x filter/.code={\pgfmathparse{#1/1e1}}]
    table[col sep=comma] {fig/residual_acf/60000.csv};
    \coordinate (legend) at (axis description cs:0.61,.60); %

  \end{axis}
  \matrix [
    draw,
    fill=white,
    fill opacity=.8,
    text opacity=1,
    matrix of nodes,
    align =left,
    row sep = 0,
    column sep = 0,
    inner sep= 3,
    anchor=south west,
    font=\footnotesize,
    mark options={solid},
    rounded corners=4pt
  ] (mymat) at (legend) {
    \draw[thick,uniblauHell!0!uniblauDunkel] (-.25,.05) -- (.25,.05);   & $N_\mathrm{S}=6000$ \\
    \draw[thick,uniblauHell!20!uniblauDunkel] (-.25,.05) -- (.25,.05);  & $N_\mathrm{S}=5000$ \\
    \draw[thick,uniblauHell!40!uniblauDunkel] (-.25,.05) -- (.25,.05);  & $N_\mathrm{S}=4000$ \\
    \draw[thick,uniblauHell!60!uniblauDunkel] (-.25,.05) -- (.25,.05);  & $N_\mathrm{S}=3000$ \\
    \draw[thick,uniblauHell!80!uniblauDunkel] (-.25,.05) -- (.25,.05);  & $N_\mathrm{S}=2000$ \\
    \draw[thick,uniblauHell!100!uniblauDunkel] (-.25,.05) -- (.25,.05); & $N_\mathrm{S}=1000$ \\
  };
\end{tikzpicture}
  \caption{Autocorrelation of residual error for different window sizes $2N_\mathrm{S}+1$ over the time lag $l$, indicating that the residual error is correlated.}
  \label{fig::residual_acf}
\end{figure}

\begin{figure}%
  \centering
  \begin{tikzpicture}
\def\numericalscale{280}
  \begin{axis}[
    width=.8*\textwidth*0.55, height=.9*\textwidth*0.55,
    xlabel={Frequency $f$ in $\mathrm{MHz}$},
    ylabel={PSD of residual $\left|S_{nn}\left( f \right)\right|$ in $\nicefrac{\mathrm{W}}{\mathrm{Hz}}$},
    tick align=outside,
    tick pos=left,
    xmin=-100, xmax=100,
    scaled x ticks=false,
    ]
    \addplot [very thick,uniblauHell!0!uniblauDunkel,mark=none, mark size=1.5, mark options={solid}, x filter/.code={\pgfmathparse{#1/1e6}}, fill=uniblauHell!0!uniblauDunkel, fill opacity=0]
    table[col sep=comma] {fig/residual_psd/60000.csv};
    
    \addplot [thick,uniblauHell!0!black,mark=none, mark size=1.5, mark options={solid}, x filter/.code={\pgfmathparse{#1/1e6}}, y filter/.code={\pgfmathparse{#1*280}}, loosely dashed]
    table[col sep=comma] {fig/residual_psd/numerical/60000.csv};

    \addplot [very thick,uniblauHell!20!uniblauDunkel,mark=none, mark size=1.5, mark options={solid}, x filter/.code={\pgfmathparse{#1/1e6}}]
    table[col sep=comma] {fig/residual_psd/50000.csv};
    \addplot [thick,uniblauHell!0!black,mark=none, mark size=1.5, mark options={solid}, x filter/.code={\pgfmathparse{#1/1e6}}, y filter/.code={\pgfmathparse{#1*333}}, loosely dashed]
    table[col sep=comma] {fig/residual_psd/numerical/50000.csv};

    \addplot [very thick,uniblauHell!40!uniblauDunkel,mark=none, mark size=1.5, mark options={solid}, x filter/.code={\pgfmathparse{#1/1e6}}]
    table[col sep=comma] {fig/residual_psd/40000.csv};
    \addplot [thick,uniblauHell!0!black,mark=none, mark size=1.5, mark options={solid}, x filter/.code={\pgfmathparse{#1/1e6}}, y filter/.code={\pgfmathparse{#1*324}}, loosely dashed]
    table[col sep=comma] {fig/residual_psd/numerical/40000.csv};

    \addplot [very thick,uniblauHell!60!uniblauDunkel,mark=none, mark size=1.5, mark options={solid}, x filter/.code={\pgfmathparse{#1/1e6}}]
    table[col sep=comma] {fig/residual_psd/30000.csv};
    \addplot [very thick,uniblauHell!0!black,mark=none, mark size=1.5, mark options={solid}, x filter/.code={\pgfmathparse{#1/1e6}}, y filter/.code={\pgfmathparse{#1*340}}, loosely dashed]
    table[col sep=comma] {fig/residual_psd/numerical/30000.csv};

    \addplot [very thick,uniblauHell!80!uniblauDunkel,mark=none, mark size=1.5, mark options={solid}, x filter/.code={\pgfmathparse{#1/1e6}}]
    table[col sep=comma] {fig/residual_psd/20000.csv};
    \addplot [very thick,uniblauHell!0!black,mark=none, mark size=1.5, mark options={solid}, x filter/.code={\pgfmathparse{#1/1e6}}, y filter/.code={\pgfmathparse{#1*355}}, loosely dashed]
    table[col sep=comma] {fig/residual_psd/numerical/20000.csv};

    \addplot [very thick,uniblauHell!100!uniblauDunkel,mark=none, mark size=1.5, mark options={solid}, x filter/.code={\pgfmathparse{#1/1e6}}]
    table[col sep=comma] {fig/residual_psd/10000.csv};
    \addplot [very thick,uniblauHell!0!black,mark=none, mark size=1.5, mark options={solid}, x filter/.code={\pgfmathparse{#1/1e6}}, y filter/.code={\pgfmathparse{#1*350}}, loosely dashed]
    table[col sep=comma] {fig/residual_psd/numerical/10000.csv};

    \coordinate (legend) at (axis description cs:0.01,.37); %

  \end{axis}
  \matrix [
    draw,
    fill=white,
    fill opacity=.8,
    text opacity=1,
    matrix of nodes,
    align =left,
    row sep = 0,
    column sep = 0,
    inner sep= 3,
    text width=40,
    anchor=south west,
    font=\footnotesize,
    mark options={solid},
    rounded corners=4pt
  ] (mymat) at (legend) {
    \draw[thick,uniblauHell!0!uniblauDunkel] (-.25,.05) -- (.25,.05);   & $N_\mathrm{S}=6000$         \\
    \draw[thick,uniblauHell!20!uniblauDunkel] (-.25,.05) -- (.25,.05);  & $N_\mathrm{S}=5000$         \\
    \draw[thick,uniblauHell!40!uniblauDunkel] (-.25,.05) -- (.25,.05);  & $N_\mathrm{S}=4000$         \\
    \draw[thick,uniblauHell!60!uniblauDunkel] (-.25,.05) -- (.25,.05);  & $N_\mathrm{S}=3000$         \\
    \draw[thick,uniblauHell!80!uniblauDunkel] (-.25,.05) -- (.25,.05);  & $N_\mathrm{S}=2000$         \\
    \draw[thick,uniblauHell!100!uniblauDunkel] (-.25,.05) -- (.25,.05); & $N_\mathrm{S}=1000$         \\
    \draw[thick,uniblauHell!0!black, loosely dashed] (-.25,.05) -- (.25,.05); & Numerical sim. \\
  };
\end{tikzpicture}
  \caption{PSD of residual error for different window sizes $2N_\mathrm{S}+1$ showing that the derived autocorrelation and PSD agree with numerical simulations of the full system.}
  \label{fig::residual_psd}
\end{figure}
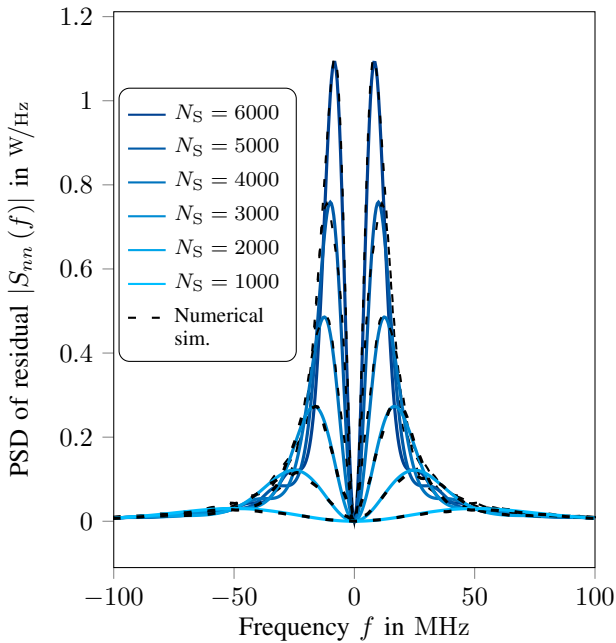

An analytical solution for the PSD is very lengthy.
Therefore, we opted to transform the analytical solution of Eq.~(\ref{eq::residual_acf}) into Frequency domain and take the absolute value as per the Wiener-Khintchine theorem to get the PSD.
Both, ACF and PSD, are shown in Fig.~\ref{fig::residual_acf} and Fig.~\ref{fig::residual_psd}, respectively.

As expected from intuition, the ACF gets wider with increasing $N$.
The more symbols are used to estimate the linear regression parameters, the more correlated the symbols become. 
The linewidth has no influence apart from a scaling factor on  both plots.
For the PSD plot in Fig.~\ref{fig::residual_psd}, we compared the pseudo-analytical result based on the ACF to direct numerical calculations as per our model.
It can be seen that they perfectly match.
The plot also emphasizes that the residual is zero-mean of the residual as it is zero at $f=0$.
With a visualization of the spectral distribution of the residual error, we can better interpret what the convolution of the signal with the residual error in frequency domain means.

It leads to spectral broadening and shows a bandpass behavior.
As a result, the bandwidth expansion needs to be considered when performing simulations of EEPN impacted communication links.
Note that, if the simulations are performed on symbol frequency, the information loss due to the bandwidth expansion of the signal is not considered. The out of band noise will be remapped to the Nyquist bandwidth (aliasing) due to the circular nature of digital signals, which can lead to overestimated performance of the simulated links if the employed mitigation makes use of this information.
 
Additionally we can see that the PSD gets narrower as $N$ increases due to the time-bandwidth properties.
Thus, the probabilities of certain frequency increases and the residual error gets more ``colored''.

\section{DSP-aware Analysis \label{sec::dsp_aware_analysis}}
As a last step, the four different terms are analyzed to study their respective impact on the receiver SNR.
The simulation parameters can be found in Tab.~\ref{tab::sim_params}.
\begin{table}[t!]
  \centering
  \caption{Parameters of the simulated system}
  \begin{tabular}{rcc}
    \toprule
    Name                               & Symbol           & Value                                           \\
    \midrule
    Symbol Rate                        & $\symbolrate$   & $100\,\mathrm{GBd}$                             \\
    Simulation frequency (CH, CDC, LO) & $\simfreq$ & $1\,\nicefrac{\mathrm{TS}}{\mathrm{s}}$         \\
    RRC roll-off                       & $\alpha$         & $0.1$                                           \\
    \midrule
    Dispersion coefficient             & $\beta_2$        & $-21.67\,\nicefrac{\mathrm{ps}^2}{\mathrm{km}}$ \\
    Fiber length                       & $\ell$           & $4000\,\mathrm{km}$                             \\
    Baseline SNR                       & $\operatorname{SNR}\bigr\vert_{\Delta\nu =0}$ & $13.7\,\mathrm{dB}$ \\
    \midrule
    No. of taps in CPR                 & $N_\mathrm{CPR}$ & $701$                                           \\
    \bottomrule
  \end{tabular}
  \label{tab::sim_params}
\end{table}
Channel simulations were performed at a frequency of $1\,\nicefrac{\mathrm{TS}}{\mathrm{s}}$ which corresponds to a bandwidth that is large enough to capture all of the relevant effects of EEPN at a maximal symbol rate $\symbolrate$ of $100\,\mathrm{GBd}$.
The power of the complex AWGN noise from the channel was chosen such that the baseline SNR for $\Delta\nu = 0$ is $13.7\,\mathrm{dB}$.
For the stated assumption of an ideal chromatic dispersion compensation (CDC), the compensation was also performed on simulation frequency $\simfreq$.
All the typical blocks in the DSP chain were performed at the respective usual rate, i.e., timing recovery using the Gardner algorithm at double the symbol rate and CPR at the symbol rate.
As a CPR scheme, IDR with appropriately chosen memory length as in \cite{Arnould2019} was used.
Due to the introduction of a transmit laser, the memory length of CPR was increased compared to \cite{Jung2024}.
The fiber was assumed to be a standard single-mode fiber with $\beta_2 = -21.67\,\nicefrac{\mathrm{ps}^2}{\mathrm{km}}$ at varying lengths.
For the simulations, a fiber length of $4000\,\mathrm{km}$ was used, if not stated otherwise, to have a noticeable penalty by EEPN.

\begin{figure}
  \centering
  \begin{tikzpicture}
  \begin{groupplot}[group style={group size=2 by 2, vertical sep=.5cm}]
    \nextgroupplot[
      tick align=inside,
      tick pos=left,
      xmin=451.0, xmax=5051.0,
      ymin=451.0, ymax=1251.0,
      axis on top,
      ylabel={CPR Avglen},
      xticklabels={,,},
      title={Penalty by \xte},
      colorbar,
      colormap={mymap}{[1pt]
          rgb(0pt)=(1.000,0.961,0.941);
          rgb(1pt)=(0.996,0.878,0.824);
          rgb(2pt)=(0.988,0.733,0.631);
          rgb(3pt)=(0.988,0.573,0.447);
          rgb(4pt)=(0.984,0.416,0.290);
          rgb(5pt)=(0.937,0.231,0.173);
          rgb(6pt)=(0.796,0.094,0.114);
          rgb(7pt)=(0.647,0.059,0.082);
          rgb(8pt)=(0.404,0.000,0.051)
        },
      point meta min=0.17755957693156788,
      point meta max=0.3375512825194402,
      width=0.25\textwidth,
      height=0.25\textwidth*0.7,
      yticklabel style={font=\tiny, xshift=.1cm},
      xticklabel style={font=\tiny},
      xlabel style={font=\tiny},
      ylabel style={font=\tiny, yshift=-.5cm},
      colorbar style={font=\tiny, ylabel={SNR penalty in dB}, ylabel style={font=\tiny,yshift=-1.1cm}, width=.2cm, xshift=-.07cm, yticklabel style={rotate=-90, yshift=-.1cm}, yticklabels={0.2,0.3}, ytick={0.2,0.3}},
      title style={font=\footnotesize, yshift=-.3cm},
    ]
    \addplot graphics [includegraphics cmd=\pgfimage,xmin=451.0, xmax=5051.0, ymin=451.0, ymax=1251.0,] { 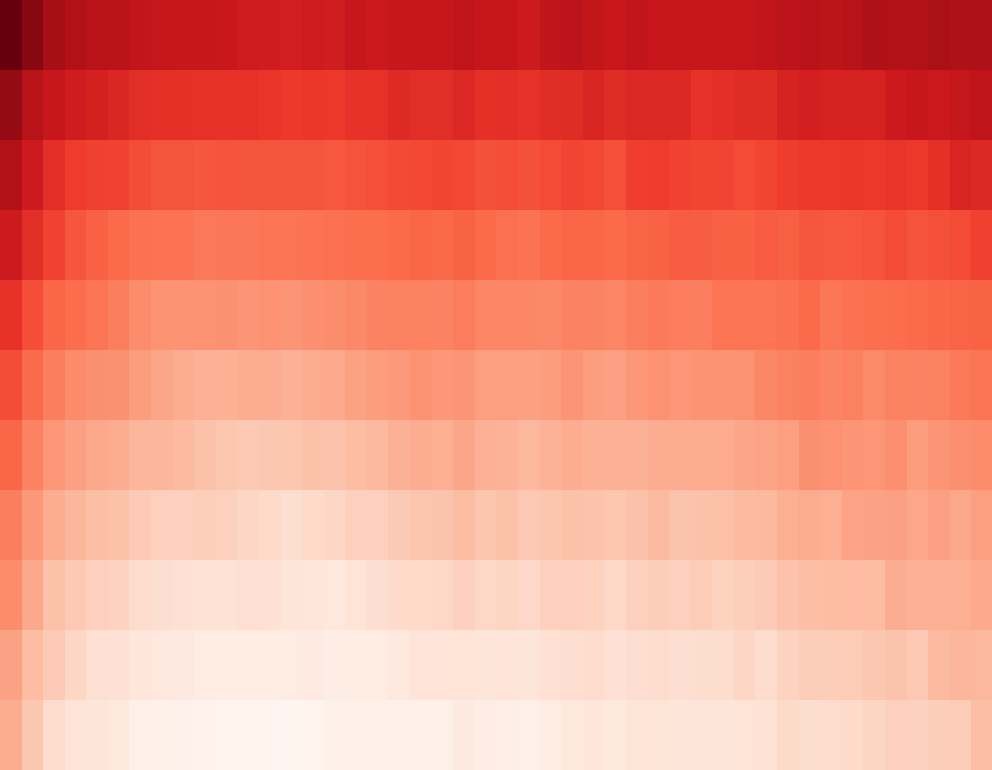 };

    \nextgroupplot[
      tick align=inside,
      tick pos=left,
      xmin=451.0, xmax=5051.0,
      ymin=451.0, ymax=1251.0,
      axis on top,
      ylabel={CPR Avglen},
      xticklabels={,,},
      title={Penalty by \nro},
      colorbar,
      colormap={mymap}{[1pt]
          rgb(0pt)=(1.000,0.961,0.941);
          rgb(1pt)=(0.996,0.878,0.824);
          rgb(2pt)=(0.988,0.733,0.631);
          rgb(3pt)=(0.988,0.573,0.447);
          rgb(4pt)=(0.984,0.416,0.290);
          rgb(5pt)=(0.937,0.231,0.173);
          rgb(6pt)=(0.796,0.094,0.114);
          rgb(7pt)=(0.647,0.059,0.082);
          rgb(8pt)=(0.404,0.000,0.051)
        },
      point meta min=0.11371883545023742,
      point meta max=0.25600698841500424,
      width=0.25\textwidth,
      height=0.25\textwidth*0.7,
      yticklabels={,,},
      ylabel={},
      yticklabel style={font=\tiny},
      xticklabel style={font=\tiny},
      xlabel style={font=\tiny, yshift=.5cm},
      ylabel style={font=\tiny, xshift=-.5cm},
      colorbar style={font=\tiny, ylabel={SNR penalty in dB}, ylabel style={font=\tiny,yshift=-1.1cm}, width=.2cm, xshift=-.07cm, yticklabel style={rotate=-90, yshift=-.1cm}, yticklabels={0.15,0.20}, ytick={0.15,0.20}},
      title style={font=\footnotesize, yshift=-.3cm},
    ]
    \addplot graphics [includegraphics cmd=\pgfimage,xmin=451.0, xmax=5051.0, ymin=451.0, ymax=1251.0,] { 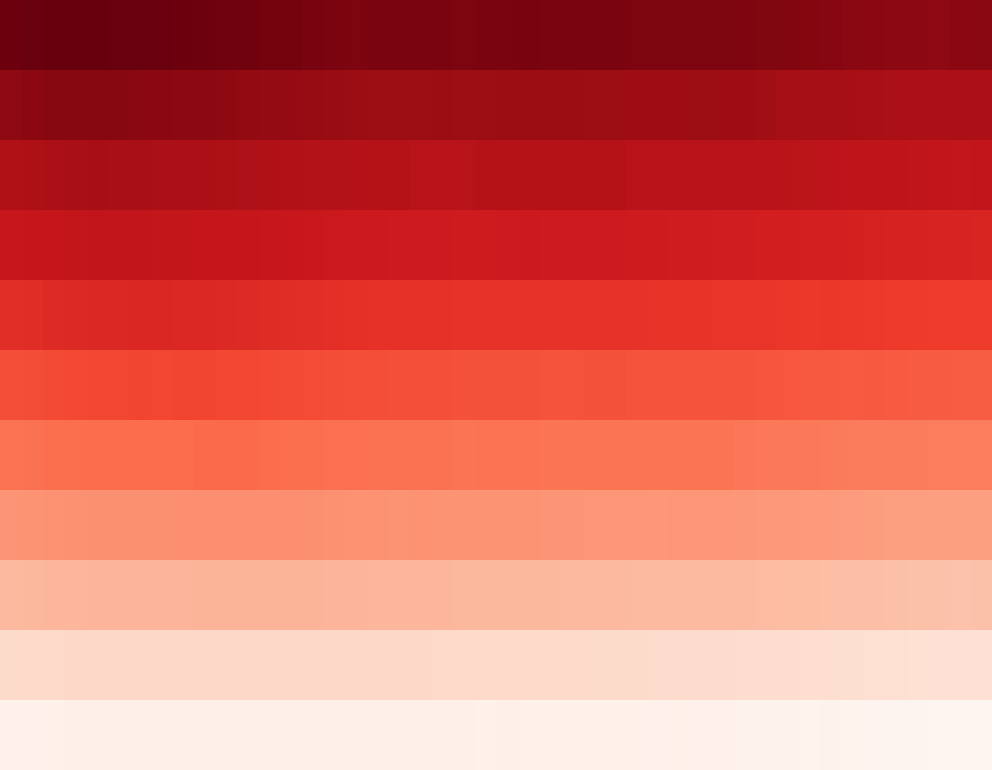 };
    \nextgroupplot[
      tick align=inside,
      tick pos=left,
      xmin=451.0, xmax=5051.0,
      ymin=451.0, ymax=1251.0,
      axis on top,
      ylabel={CPR Avglen},
      xlabel={Gardner Avglen},
      title={Penalty by \nrr},
      colorbar,
      colormap={mymap}{[1pt]
          rgb(0pt)=(1.000,0.961,0.941);
          rgb(1pt)=(0.996,0.878,0.824);
          rgb(2pt)=(0.988,0.733,0.631);
          rgb(3pt)=(0.988,0.573,0.447);
          rgb(4pt)=(0.984,0.416,0.290);
          rgb(5pt)=(0.937,0.231,0.173);
          rgb(6pt)=(0.796,0.094,0.114);
          rgb(7pt)=(0.647,0.059,0.082);
          rgb(8pt)=(0.404,0.000,0.051)
        },
      point meta min=0.22772028952710954,
      point meta max=0.23887216462181016,
      width=0.25\textwidth,
      height=0.25\textwidth*0.7,
      ylabel={CPR Avglen},
      yticklabel style={font=\tiny, xshift=.1cm},
      xticklabel style={font=\tiny},
      xlabel style={font=\tiny, yshift=.2cm},
      ylabel style={font=\tiny, yshift=-.5cm},
      colorbar style={font=\tiny, ylabel={SNR penalty in dB}, ylabel style={font=\tiny,yshift=-1.1cm}, width=.2cm, xshift=-.07cm, yticklabel style={rotate=-90, yshift=-.1cm}, yticklabels={0.23,0.235}, ytick={0.23,0.235}},
      title style={font=\footnotesize, yshift=-.3cm},
    ]
    \addplot graphics [includegraphics cmd=\pgfimage,xmin=451.0, xmax=5051.0, ymin=451.0, ymax=1251.0,] { 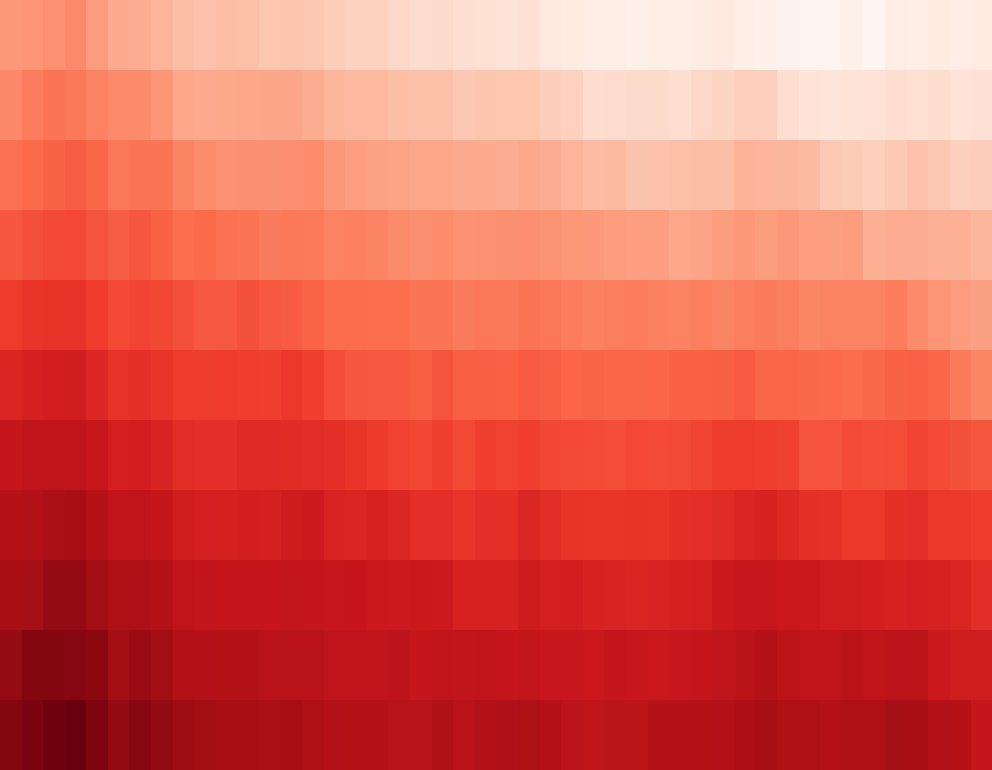 };
    \nextgroupplot[
      tick align=inside,
      tick pos=left,
      xmin=451.0, xmax=5051.0,
      ymin=451.0, ymax=1251.0,
      axis on top,
      ylabel={CPR Avglen},
      xlabel={Gardner Avglen},
      title={Penalty by \nxr},
      colorbar,
      colormap={mymap}{[1pt]
          rgb(0pt)=(1.000,0.961,0.941);
          rgb(1pt)=(0.996,0.878,0.824);
          rgb(2pt)=(0.988,0.733,0.631);
          rgb(3pt)=(0.988,0.573,0.447);
          rgb(4pt)=(0.984,0.416,0.290);
          rgb(5pt)=(0.937,0.231,0.173);
          rgb(6pt)=(0.796,0.094,0.114);
          rgb(7pt)=(0.647,0.059,0.082);
          rgb(8pt)=(0.404,0.000,0.051)
        },
      point meta min=0.0011041691043391921,
      point meta max=0.0012534682378664286,
      width=0.25\textwidth,
      height=0.25\textwidth*0.7,
      yticklabels={,,},
      ylabel={},
      yticklabel style={font=\tiny},
      xticklabel style={font=\tiny},
      xlabel style={font=\tiny, yshift=.2cm},
      ylabel style={font=\tiny},
      colorbar style={font=\tiny, width=.2cm, xshift=-.07cm, ylabel={SNR penalty in dB}, ylabel style={font=\tiny,yshift=-1.1cm}, yticklabel style={rotate=-90, yshift=-.1cm, scaled y ticks=false}, ytick={0.00115, 0.00125}, yticklabels={0.00115,0.00125}},
      title style={font=\footnotesize, yshift=-.3cm},
    ]
    \addplot graphics [includegraphics cmd=\pgfimage,xmin=451.0, xmax=5051.0, ymin=451.0, ymax=1251.0,] { 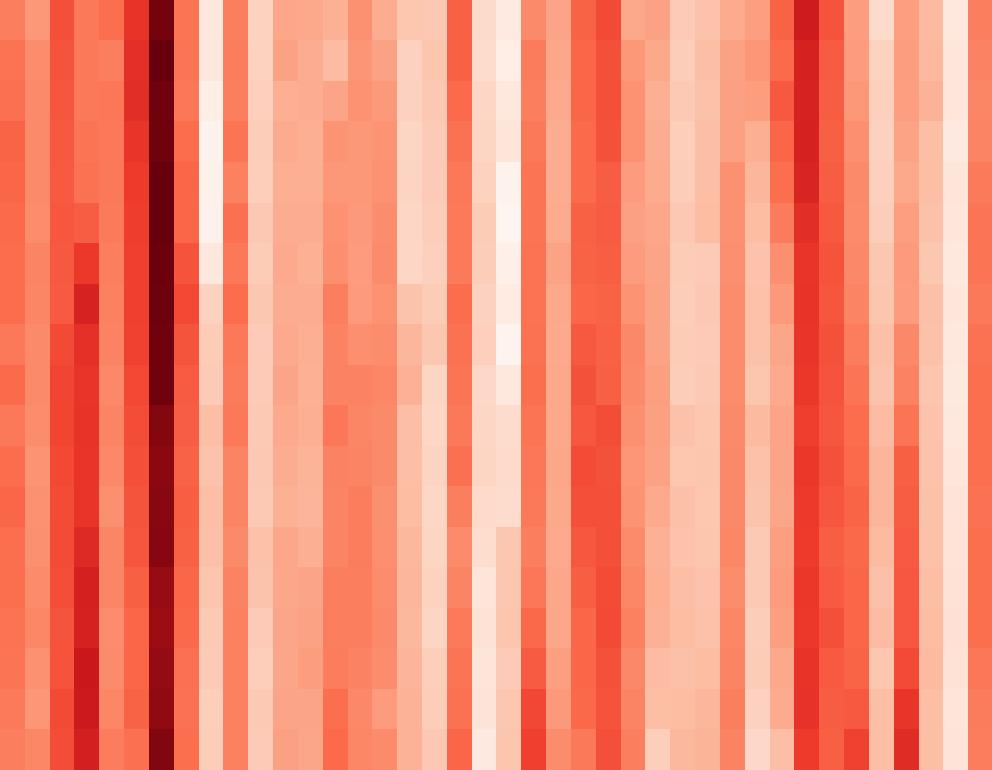 };
  \end{groupplot}
\end{tikzpicture}
  \caption{Heatmaps of link budget penalty in $\mathrm{dB}$ contributed by each term of the EEPN model for different averaging lengths in CPR and TR using a dispersive channel with AWGN.}
  \label{fig::penalty_maps}
\end{figure}

The impact of each term is then analyzed individually giving the net penalty induced by the respective term.
Due to the fact that only the timing error term $\xte$ contains unambiguous information about the transmitted signal, the penalties of all other terms were calculated in combination with this term.
By checking the different combinations and calculating the impact of each term in every scenario, we can state the net penalty by each term reliably.
The penalty is also dependent on Gardner TR and CPR. Hence, the impact has to be considered  for different averaging lengths of the two.
The results are presented in Fig.~\ref{fig::penalty_maps}.
The different averaging lengths are given on the $x$-axis and the $y$-axis for CPR and TR, respectively.
A darker color indicates a higher penalty on average.
However, each plot uses a different scaling for the heatmap, and thus, they cannot be compared directly.

Starting with the first term, the timing error term $\xte$, it can be seen that the averaging length in TR has an impact on the penalty which is to be expected.
Additionally, the penalty increases with longer averaging lengths in CPR as in \cite{Arnould2019}.
In terms of the quantitatively induced penalty, the timing error term is the most dominant one.

For the rotation term, it can be seen that there is no dependency on the timing recovery, but rather, only on the CPR.
This strong dependency on the CPR is in line with the expectations resulting from Fig.~\ref{fig::EEPN_terms} where it was shown that this term induces further rotations stemming from the transmit laser.
The relative penalty is lower than for the first term -- but still significant.

The receiver residual term has no dependency on either CPR or TR.
The induced penalty stays pretty much constant for all averaging lengths and DSP blocks
Hence, this term stays fairly unchanged after the classical DSP pipeline and gives a penalty of roughly $0.2\,\mathrm{dB}$ at $150\,\mathrm{kHz}$.

Lastly, the cross residual term is negligible in terms of its impact compared to the others.
It is, similar to the receiver residual term, constant over all averaging lengths and therefore is not influenced by the DSP blocks.
Interestingly, this term achieves values larger than zero indicating a performance gain which can be explained by the small overall impact and change to the symbols. %
Hence, the penalty is not directly comparable to the other terms.

\begin{figure}
  \centering
  \begin{tikzpicture}
  \def\xminfirstterm{150000.0}
  \def\xmaxfirstterm{1000000.0}
  \def\yminfirstterm{-0.16686781087361452}
  \def\ymaxfirstterm{3.547809289328112}
  \def\xminsecondterm{150000.0}
  \def\xmaxsecondterm{1000000.0}
  \def\yminsecondterm{0.05318506958221027}
  \def\ymaxsecondterm{0.9740122705005648}
  \def\xminthirdterm{150000.0}
  \def\xmaxthirdterm{1000000.0}
  \def\yminthirdterm{0.0567292058615967}
  \def\ymaxthirdterm{1.5907344783917332}
  \def\xminfourthterm{150000.0}
  \def\xmaxfourthterm{1000000.0}
  \def\yminfourthterm{-0.0031712983687013714}
  \def\ymaxfourthterm{0.05758017903398226}
  \begin{groupplot}[group style={group size=2 by 2, vertical sep=.6cm, horizontal sep=.7cm}]
    \nextgroupplot[
      tick align=outside,
      tick pos=left,
      xmin=\xminfirstterm,
      xmax=\xmaxfirstterm,
      ymin=\yminfirstterm,
      ymax=\ymaxfirstterm,
      width=.28\textwidth,
      height=.28\textwidth*0.625,
      xticklabel style={font=\tiny},
      yticklabel style={font=\tiny, xshift=.05cm},
      xlabel={},
      xlabel style={font=\tiny, yshift=.3cm},
      ylabel={Penalty in $\mathrm{dB}$},
      ylabel style={font=\tiny, yshift=-.6cm},
      title={Penalties for \xte},
      title style={font=\footnotesize, yshift=-.2cm},
      xtick = {150000,300000,500000,1000000},
      major tick length=0.1cm,
      xticklabels = {,,},
      scaled x ticks = false,
    ]
    \addplot graphics [includegraphics cmd=\pgfimage,xmin=\xminfirstterm,xmax=\xmaxfirstterm,ymin=\yminfirstterm,ymax=\ymaxfirstterm] {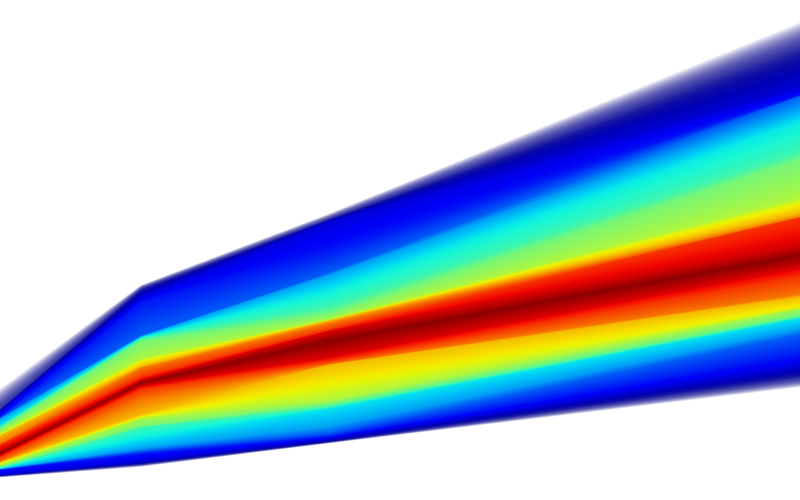};
    \nextgroupplot[
      tick align=outside,
      tick pos=left,
      xmin=\xminsecondterm,
      xmax=\xmaxsecondterm,
      ymin=\yminsecondterm,
      ymax=\ymaxsecondterm,
      width=.28\textwidth,
      height=.28\textwidth*0.625,
      xticklabel style={font=\tiny},
      yticklabel style={font=\tiny, xshift=.05cm},
      xlabel={},
      xlabel style={font=\tiny, yshift=.3cm},
      ylabel={},
      ylabel style={font=\tiny, yshift=-.6cm},
      title={Penalties for \nro},
      title style={font=\footnotesize, yshift=-.2cm},
      xtick = {150000,300000,500000,1000000},
      major tick length=0.1cm,
      xticklabels = {,,,},
      scaled x ticks = false,
    ]
    \addplot graphics [includegraphics cmd=\pgfimage,xmin=\xminsecondterm,xmax=\xmaxsecondterm,ymin=\yminsecondterm,ymax=\ymaxsecondterm] {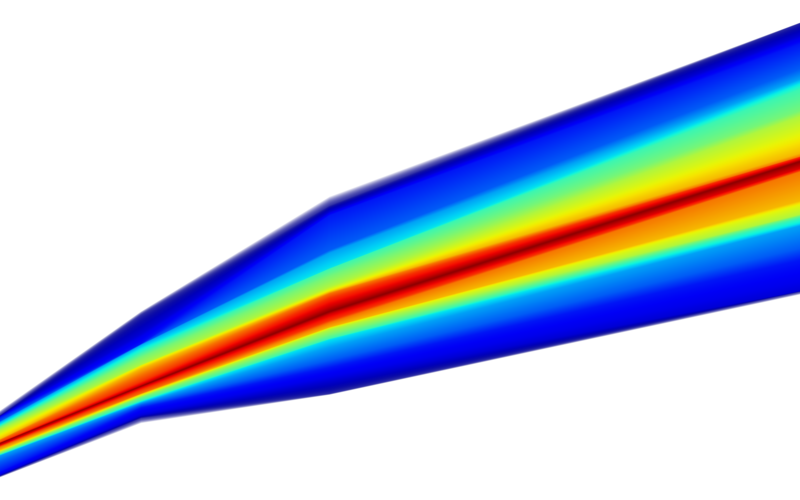};
    \nextgroupplot[
      tick align=outside,
      tick pos=left,
      xmin=\xminthirdterm,
      xmax=\xmaxthirdterm,
      ymin=\yminthirdterm,
      ymax=\ymaxthirdterm,
      width=.28\textwidth,
      height=.28\textwidth*0.625,
      xticklabel style={font=\tiny},
      yticklabel style={font=\tiny, xshift=.05cm},
      xlabel={Linewidth in $\mathrm{kHz}$},
      xlabel style={font=\tiny, yshift=.3cm},
      ylabel={Penalty in $\mathrm{dB}$},
      ylabel style={font=\tiny, yshift=-.6cm},
      title={Penalties for \nrr},
      title style={font=\footnotesize, yshift=-.2cm},
      xtick = {150000,300000,500000,1000000},
      major tick length=0.1cm,
      xticklabels = {150,300,500,1000},
      scaled x ticks = false,
    ]
    \addplot graphics [includegraphics cmd=\pgfimage,xmin=\xminthirdterm,xmax=\xmaxthirdterm,ymin=\yminthirdterm,ymax=\ymaxthirdterm] {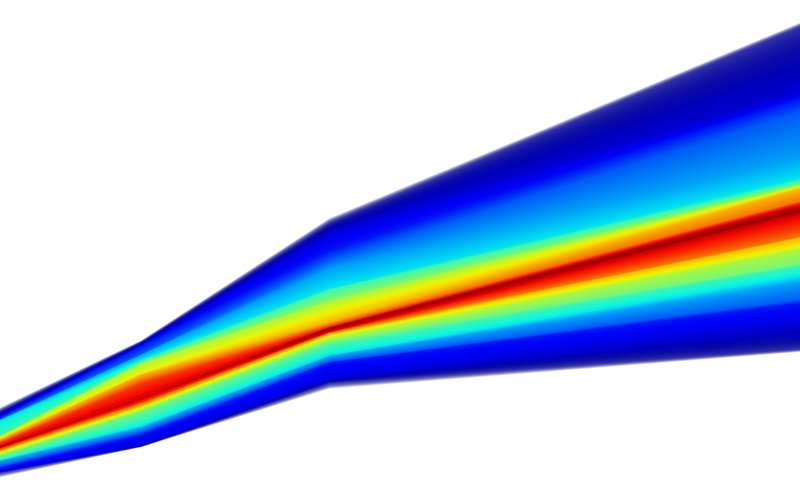};
    \nextgroupplot[
      tick align=outside,
      tick pos=left,
      xmin=\xminfourthterm,
      xmax=\xmaxfourthterm,
      ymin=\yminfourthterm,
      ymax=\ymaxfourthterm,
      width=.28\textwidth,
      height=.28\textwidth*0.625,
      xticklabel style={font=\tiny},
      yticklabel style={font=\tiny, xshift=.05cm},
      xlabel={Linewidth in $\mathrm{kHz}$},
      xlabel style={font=\tiny, yshift=.3cm},
      ylabel={},
      ylabel style={font=\tiny, yshift=-.6cm},
      ytick={0,0.02,0.04},
      yticklabels={0,0.02,0.04},
      title={Penalties for \nxr},
      title style={font=\footnotesize, yshift=-.2cm},
      xtick = {150000,300000,500000,1000000},
      major tick length=0.1cm,
      xticklabels = {150,300,500,1000},
      scaled x ticks = false,
      scaled y ticks = false,
    ]
    \addplot graphics [includegraphics cmd=\pgfimage,xmin=\xminfourthterm,xmax=\xmaxfourthterm,ymin=\yminfourthterm,ymax=\ymaxfourthterm] {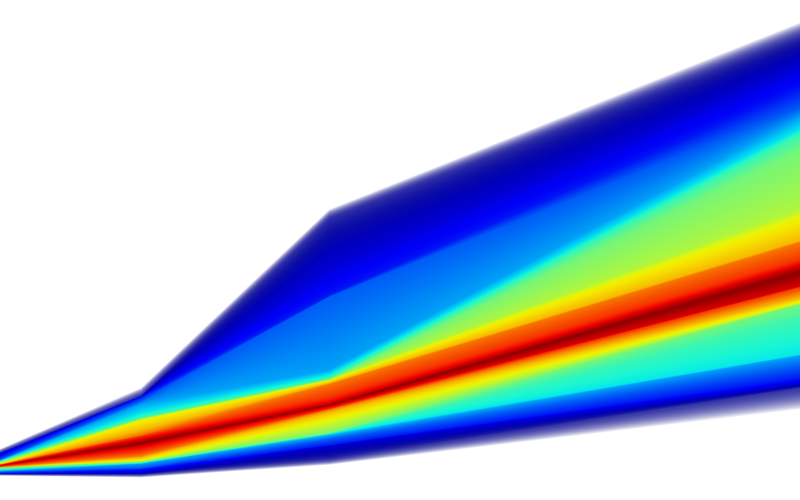};
  \end{groupplot}
\end{tikzpicture}
  \vspace*{-.8cm}
  \caption{Link budget penalty in $\mathrm{dB}$ per impairment term of the proposed model, versus linewidth.}
  \label{fig::penalty_linewidth}
\end{figure}

Up to this point, the analysis was performed for a linewidth of $150\,\mathrm{kHz}$.
Fig.~\ref{fig::penalty_linewidth} shows the behavior for different linewidths.
The penalty distribution for each term over different linewidths is plotted.
The averaging lengths of the CPR and TR were fixed to 701 and 1501, respectively.

It can be seen that the average penalty for $\xte$, $\nro$, and $\nrr$ show similar behavior.
The penalty does not increase linearly with the linewidth but rather shows a more logarithmic behavior.

It is important to note that this is the penalty after TR and CPR.
Otherwise the penalties would show a linear response.

The observed behavior of the penalty can also be found in \cite{Jung2024}.
There, TR was able to reduce more penalty the stronger EEPN got.

Only $\nxr$  shows a different behavior and seems to exponentially increase with the linewidth.
However, the penalty is still negligible compared to the other terms.

Wrapping up the results section, we conclude that by formulating the phase noise in the proposed way, that lead to constructive guidelines for RX DSP pipeline design.
It also enables a novel analysis of EEPN penalty with respect to the influence of different DSP blocks.
TR only influences one term of EEPN significantly.
This motivates the gained interest in TR with respect to EEPN in research in the last years \cite{Martins2024,Qiu2024}.
Evidently, the receiver and cross residual terms remain largely unaffected by the DSP blocks under consideration.

\section{Conclusion \label{sec::Conclusion}}
In this paper, we propose a novel model to describe phase noise and consequently EEPN.
The model enables to split the received EEPN affected signal into four main contributors: timing error $\xte$, rotation $\nro$, receiver residual $\nrr$, and cross residual $\nxr$.
For each term, a stochastic model allows to understand its behavior and the impact on the original signal.

Furthermore, we can estimate the impact of classical coherent DSP blocks on the signal with the model and describe the resulting signal after applying them.

In future work, the proposed model can also be combined with, e.g., the filters proposed in \cite{Xu2024} to better match a real world laser.

\IEEEtriggeratref{9}
\bibliographystyle{IEEEtran}
\bibliography{refs}
\clearpage

\onecolumn
\appendix
\section{Appendix}
\numberwithin{equation}{subsection}
\subsection{Variance of the residual noise}

\begin{align}
  \nonumber
  \mathrm{Var}\left[n_k^2\right]  = & \mathrm{E}\left[n_k^2\right] = \mathrm{E}\left[\left(\phi_k  -\frac{1}{2N+1}\sum_{i=-N}^{N}\phi_{k+i}\right)^2\right]                                                                                             \\
  =                                 & \mathrm{Var}\left(\phi_0 + \frac{1}{2N+1}\sum\limits_{i=-N}^{N}\Delta\phi_i\right)                                                                                                                        \\
  \begin{split}
    = & \frac{1}{\left(2N+1\right)^2}\sum\limits_{i=-N}^{N}i\mathrm{Var}\left(\Delta\phi_i\right)                                                                                                                         \\
    & + \frac{2}{\left(2N+1\right)^2}\left[\sum\limits_{i=-N}^{-1}\sum\limits_{j=-N}^{-i}j\mathrm{Var}\left(\Delta \phi_j\right)+\sum\limits_{i=1}^{N}\sum\limits_{j=i}^{N}j\mathrm{Var}\left(\Delta \phi_j\right)\right] \\
  \end{split} \\
  =                                 & \frac{N\left(N+1\right)}{\left(2N+1\right)^2}\frac{2\pi\lolw}{\simfreq} + \frac{4}{\left(2N+1\right)^2}\sum\limits_{i=1}^{N}\sum\limits_{j=1}^{N-i}j\frac{2\pi\lolw}{\simfreq}        \\
  \begin{split}
    =& \frac{N\left(N+1\right)}{\left(2N+1\right)^2}\frac{2\pi\lolw}{\simfreq}                                                                                                                          \\
    & + \frac{4}{\left(2N+1\right)^2}\frac{2\pi\lolw}{\simfreq}\sum\limits_{i=1}^{N}\frac{\left(N-i\right)\left(N-i+1\right)}{2}                                                                         \\
  \end{split}                      \\
  \begin{split}
    =& \frac{N\left(N+1\right)}{\left(2N+1\right)^2}\frac{2\pi\lolw}{\simfreq}                                                                                                                       \\
    & + \frac{2}{\left(2N+1\right)^2}\frac{2\pi\lolw}{\simfreq}\sum\limits_{i=1}^{N}N^2-2iN+N+i^2-i                                                                                                   \\
  \end{split}                         \\
  \begin{split}
    =& \frac{N\left(N+1\right)}{\left(2N+1\right)^2}\frac{2\pi\lolw}{\simfreq}                                                                                                                       \\
    & + \frac{2}{\left(2N+1\right)^2}\frac{2\pi\lolw}{\simfreq}\left(N^3+N^2 - 2N\frac{N\left(N+1\right)}{2}-\frac{N\left(N+1\right)}{2}+\frac{N\left(N+1\right)\left(2N+1\right)}{6}\right)          \\
  \end{split}                          \\
  \begin{split}
    =& \frac{N\left(N+1\right)}{\left(2N+1\right)^2}\frac{2\pi\lolw}{\simfreq}                                                                                                                       \\
    & + \frac{2}{\left(2N+1\right)^2}\frac{2\pi\lolw}{\simfreq}\frac{N^3-N}{3}                                                                                                                        \\
  \end{split}                         \\
                                    & = \frac{2\pi\lolw}{\simfreq\left(2N+1\right)^2}\left(N\left(N+1\right)+\frac{2N^3-2N}{3}\right)                                                                                                 \\
                                    & = \frac{4\pi\Delta \nu}{\simfreq\left(2N+1\right)^2}\left(\frac{2}{3}N^3 + N^2+\frac{1}{3}N\right)                                                                                                  %
\end{align}

\newpage
\subsection{ Autocorrelation of the residual noise\label{subsec::autocorrelation_derivation}}\label{subsec:autocor}
\begin{align}
  R_{\phi\phi}\left( l \right) =  \mathrm{E}\left[ \phi_k \phi_{k-l} \right]
\end{align}
With
\begin{align}
  \phi_k = \sum\limits_{i=1}^{k} \Delta\phi_i + \varphi - \frac{1}{2N+1}\sum\limits_{i=-N}^{N}\phi_{k+i}
\end{align}
\begin{align}
  R_{\phi\phi}\left( l \right) & =  \mathrm{E}\left[ \left( \sum\limits_{i=1}^{k} \Delta\phi_i + \varphi - \frac{1}{2N+1}\sum\limits_{i=-N}^{N}\Delta\phi_{k+i} \right) \left( \sum\limits_{i=1}^{k+l} \Delta\phi_i + \varphi - \frac{1}{2N+1}\sum\limits_{i=-N}^{N}\phi_{k+l+i} \right) \right] \\
  \begin{split}
    =& \mathrm{E}\left[ \sum_{i=1}^{k}\Delta\phi_i \sum\limits_{i=1}^{k+l}\Delta\phi_i \right] + \mathrm{E}\left[\sum_{i=1}^{k}\Delta\phi_i \varphi \right]\\
    &+ \mathrm{E}\left[ \left( \sum_{i=1}^{k}\Delta\phi_i \right) \left( -\frac{1}{2N+1}\sum\limits_{i=-N}^{N} \phi_{k+l+i} \right) \right]\\
    &+ \mathrm{E}\left[ \varphi \sum\limits_{i=1}^{k+l}\Delta\phi_i \right] + \mathrm{E}\left[ \varphi \varphi \right]\\
    & + \mathrm{E}\left[ \varphi \left( -\frac{1}{2N+1}\sum\limits_{i=-N}^{N} \phi_{k+l+i} \right) \right]\\
    & + \mathrm{E}\left[ \left( -\frac{1}{2N+1}\sum\limits_{i=-N}^{N} \phi_{k+i} \right) \sum\limits_{i=1}^{k+l} \Delta\phi_i \right]\\
    & + \mathrm{E}\left[ \left(- \frac{1}{2N+1}\sum\limits_{i=-N}^{N} \phi_{k+i} \right) \varphi \right]\\
    & + \mathrm{E}\left[  \frac{1}{\left(2N+1\right)^2}\left(\sum\limits_{i=-N}^{N} \phi_{k+i} \right) \left( \sum\limits_{i=-N}^{N} \phi_{k+l+i} \right) \right]
  \end{split}                                                                                                                             \\
  \begin{split}
    =& \mathrm{E}\left[ \sum_{i=1}^{k}\Delta\phi_i \sum\limits_{i=1}^{k+l}\Delta\phi_i \right] \\
    &+ \mathrm{E}\left[ \left( \sum_{i=1}^{k}\Delta\phi_i \right) \left( -\frac{1}{2N+1}\sum\limits_{i=-N}^{N} \phi_{k+l+i} \right) \right]\\
    & + \mathrm{E}\left[ \left( -\frac{1}{2N+1}\sum\limits_{i=-N}^{N} \phi_{k+i} \right) \sum\limits_{i=1}^{k+l} \Delta\phi_i \right]\\
    & + \mathrm{E}\left[  \frac{1}{\left(2N+1\right)^2}\left(\sum\limits_{i=-N}^{N} \phi_{k+i} \right) \left( \sum\limits_{i=-N}^{N} \phi_{k+l+i} \right) \right]
  \end{split}                                                                                                                                                                                          \\
  \begin{split}
    =& \mathrm{E}\left[ \sum_{i=1}^{k}\Delta\phi_i \sum\limits_{i=1}^{k+l}\Delta\phi_i \right] \\
    &+ \mathrm{E}\left[ \left( \sum_{i=1}^{k}\Delta\phi_i \right) \left( -\frac{1}{2N+1}\sum\limits_{i=0}^{2N} \sum\limits\phi_{k+l+i} \right) \right]\\
    & + \mathrm{E}\left[ \left( -\frac{1}{2N+1}\sum\limits_{i=-N}^{N} \phi_{k+i} \right) \sum\limits_{i=1}^{k+l} \Delta\phi_i \right]\\
    & + \mathrm{E}\left[  \frac{1}{\left(2N+1\right)^2}\left(\sum\limits_{i=-N}^{N} \phi_{k+i} \right) \left( \sum\limits_{i=-N}^{N} \phi_{k+l+i} \right) \right]
  \end{split}
\end{align}
Calculating the different terms for $l\geq0$ as $R_{\phi\phi}\left( l \right) = R_{\phi\phi}\left( -l \right)$
\subsubsection*{First part}
\renewcommand{\qedsymbol}{\ensuremath{\square}}
\begin{align}
  \mathrm{E}\left[ \sum_{i=1}^{k}\Delta\phi_i \sum\limits_{i=1}^{k+l}\Delta\phi_i \right] & = \min \left\{ k,k+l \right\}\sigma^2
\end{align}%
\subsubsection*{Second part}
\begin{align}
    & \mathrm{E}\left[ \left( \sum_{i=1}^{k}\Delta\phi_i \right) \left( -\frac{1}{2N+1}\sum\limits_{i=-N}^{N} \phi_{k+l+i} \right) \right]                                                                                                                                                                  \\
  = & \mathrm{E}\left[ \left( \sum_{i=1}^{k}\Delta\phi_i \right) \left(-\left(\sum\limits_{i=1}^{k+l-N-1}\Delta\phi_i + \frac{1}{2N+1}\sum\limits_{i=0}^{2N}\sum\limits_{j=0}^{i} \Delta\phi_{k+l-N+j}\right) \right)\right]                                                                                \\
  = & \mathrm{E}\left[ \left( \sum_{i=1}^{k}\Delta\phi_i \right) \left(-\left(\sum\limits_{i=1}^{k+l-N-1}\Delta\phi_i + \frac{1}{2N+1}\sum\limits_{i=0}^{2N}\left(2N+1-i  \right) \Delta\phi_{k+l-N+i}\right)\right) \right]                                                                                \\
  = & -\mathrm{E}\left[ \left(\sum\limits_{i=1}^{k} \Delta\phi_i\right)\left( \sum\limits_{i=1}^{k+l-N-1} \Delta\phi_i\right) \right] - \frac{1}{2N+1}\mathrm{E}\left[ \left( \sum\limits_{i=1}^k\Delta\phi_i \right)\left( \sum\limits_{i=0}^{2N}\left( 2N+1-i \right)\Delta\phi_{k+l-N+i} \right) \right] \\
  = & -\sigma^{2}\min\left\{ k,k+l-N-1 \right\} - \frac{1}{2N+1}\begin{cases}
                                                                  \mathrm{E}\left[ \left(\sum\limits_{i=k+l-N}^k \Delta\phi_i\right)\left( \sum\limits_{i=0}^{N-l} \left( 2N+1-i \right) \Delta\phi_{k+l-N+i} \right)\right], & 0\leq l \leq N \\
                                                                  0,                                                                                                                                                          & l >N
                                                                \end{cases}                                                                \\
  = & -\sigma^{2}\min\left\{ k,k+l-N-1 \right\} - \frac{1}{2N+1}\begin{cases}
                                                                  \mathrm{E}\left[ \left(\sum\limits_{i=l-2N}^{l-N} \Delta\phi_{k+i}\right)\left( \sum\limits_{i=l-N}^{0} \left( N+1+l-i \right) \Delta\phi_{k+i} \right)\right], & 0\leq l \leq N \\
                                                                  0,                                                                                                                                                              & l >N
                                                                \end{cases}                                                            \\
  = & -\sigma^{2}\min\left\{ k,k+l-N-1 \right\} - \frac{\sigma^{2}}{2N+1}\begin{cases}
                                                                           \sum\limits_{i=l-N}^{0}  N+1+l-i  , & 0\leq l \leq N \\
                                                                           0                                   & l >N
                                                                         \end{cases}                                                                                                                                                                               \\
  = & -\sigma^{2}\min\left\{ k,k+l-N-1 \right\} - \frac{\sigma^{2}}{2N+1}\begin{cases}
                                                                           \left( N-l+1 \right)N + N-l+1 + l\left( N-l+1 \right) + \sum\limits_{i=0}^{N-l} i  , & 0\leq l \leq N \\
                                                                           0,                                                                                   & l >N
                                                                         \end{cases}                                                                                                                              \\
  = & -\sigma^{2}\min\left\{ k,k+l-N-1 \right\} - \frac{\sigma^{2}}{2N+1}\begin{cases}
                                                                           N^2 - Nl + N + N-l+1 + Nl-l^2+l + \frac{\left(N-l\right)\left( N-l+1 \right)}{2}, & 0\leq l \leq N \\
                                                                           0,                                                                                & l >N
                                                                         \end{cases}                                                                                                                                 \\
  = & -\sigma^{2}\min\left\{ k,k+l-N-1 \right\} - \frac{\sigma^{2}}{2N+1}\begin{cases}
                                                                           N^2 + 2N + 1 - l^2 + \frac{N^2-Nl+N-Nl+l^2-l}{2}, & 0\leq l \leq N \\
                                                                           0,                                                & l >N
                                                                         \end{cases}                                                                                                                                                                 \\
  = & -\sigma^{2}\min\left\{ k,k+l-N-1 \right\} - \frac{\sigma^{2}}{2N+1}\begin{cases}
                                                                           N^2 + 2N + 1 - l^2 + \frac{N^2}{2} - Nl +\frac{N}{2} + \frac{l^2}{2} - \frac{l}{2}, & 0\leq l \leq N \\
                                                                           0,                                                                                  & l >N
                                                                         \end{cases}                                                                                                                               \\
  = & -\sigma^{2}\min\left\{ k,k+l-N-1 \right\} - \frac{\sigma^{2}}{2N+1}\begin{cases}
                                                                           \frac{3N^2}{2} + \frac{5N}{2} + 1 - \frac{l^2}{2} -Nl - \frac{l}{2}, & 0\leq l \leq N \\
                                                                           0,                                                                   & l >N
                                                                         \end{cases}
\end{align}
\newpage
\subsubsection*{Third part}
\begin{align}
    & \mathrm{E}\left[ -\frac{1}{2N+1}\left(\sum\limits_{i=-N}^{N}\phi_k+i\right)\left( \sum\limits_{i=1}^{k+l}\Delta\phi_i \right) \right]                                                                                                                                                                \\
  = & -\frac{1}{2N+1} \mathrm{E}\left[ \left(\left( 2N+1 \right)\sum\limits_{i=1}^{k-N-1}\Delta\phi_i + \sum\limits_{i=0}^{2N}\sum_{j=0}^{i}\Delta\phi_{k-N+j}\right)\left( \sum\limits_{i=1}^{k+l}\Delta\phi_i \right) \right]                                                                            \\
  = & - \mathrm{E}\left[ \left(\sum\limits_{i=1}^{k-N-1}\Delta\phi_i\right)\left( \sum\limits_{i=1}^{k+l}\Delta\phi_i \right)\right]  -\frac{1}{2N+1} \mathrm{E}\left[\left(\sum\limits_{i=0}^{2N}\left( 2N+1-i \right)\Delta\phi_{k-N+i}\right)\left( \sum\limits_{i=1}^{k+l}\Delta\phi_i \right) \right] \\
  = & - \sigma^{2}\min\left\{ k-N-1,k+l \right\}  -\frac{1}{2N+1}\begin{cases}
                                                                   \mathrm{E}\left[\left(\sum\limits_{i=0}^{N+l}\left( 2N+1-i \right)\Delta\phi_{k-N+i}\right)\left( \sum\limits_{i=k-N}^{k+l}\Delta\phi_i \right) \right], & 0\leq l  <N \\
                                                                   \mathrm{E}\left[\left(\sum\limits_{i=0}^{2N}\left( 2N+1-i \right)\Delta\phi_{k-N+i}\right)\left( \sum\limits_{i=k-N}^{k+N}\Delta\phi_i \right) \right],  & l\geq N
                                                                 \end{cases}                                                                               \\
  = & - \sigma^{2}\min\left\{ k-N-1,k+l \right\}  -\frac{1}{2N+1}\begin{cases}
                                                                   \mathrm{E}\left[\left(\sum\limits_{i=0}^{N+l}\left( 2N+1-i \right)\Delta\phi_{k-N+i}\right)\left( \sum\limits_{i=0}^{N+l}\Delta\phi_{k-N+i} \right) \right], & 0\leq l  <N \\
                                                                   \mathrm{E}\left[\left(\sum\limits_{i=0}^{2N}\left( 2N+1-i \right)\Delta\phi_{k-N+i}\right)\left( \sum\limits_{i=0}^{2N}\Delta\phi_{k-N+i} \right) \right],   & l\geq N
                                                                 \end{cases}                                                                              \\
  = & - \sigma^{2}\min\left\{ k-N-1,k+l \right\}  -\frac{\sigma^{2}}{2N+1}\begin{cases}
                                                                            \sum\limits_{i=0}^{N+l} 2N+1-i, & 0\leq l  <N \\
                                                                            \sum\limits_{i=0}^{2N} 2N+1-i , & l\geq N
                                                                          \end{cases}                                                                                                                                                                                    \\
  = & - \sigma^{2}\min\left\{ k-N-1,k+l \right\}  -\frac{\sigma^{2}}{2N+1}\begin{cases}
                                                                            2N\left( N+l+1 \right) + N+l+1  - \sum\limits_{i=0}^{N+l} i, & 0\leq l  <N \\
                                                                            2N\left( 2N+1 \right) + 2N + 1 - \sum\limits_{i=0}^{2N} i ,  & l\geq N
                                                                          \end{cases}                                                                                                                                                       \\
  = & - \sigma^{2}\min\left\{ k-N-1,k+l \right\}  -\frac{\sigma^{2}}{2N+1}\begin{cases}
                                                                            2N^2 + N + 2Nl + l +2N + 1 - \frac{\left( N+l \right)\left( N+l+1 \right)}{2}, & 0\leq l  <N \\
                                                                            4N^2 + 4N + 1 -  \frac{2N\left( 2N+1 \right)}{2},                              & l\geq N
                                                                          \end{cases}                                                                                                                                     \\
  = & - \sigma^{2}\min\left\{ k-N-1,k+l \right\}  -\frac{\sigma^{2}}{2N+1}\begin{cases}
                                                                            2N^2 + 3N + 2Nl + l + 1 - \frac{N^2 + Nl + N + Nl + l^2 + l}{2}, & 0\leq l  <N \\
                                                                            4N^2 + 4N + 1 - 2N^2 - N,                                        & l\geq N
                                                                          \end{cases}                                                                                                                                                   \\
  = & - \sigma^{2}\min\left\{ k-N-1,k+l \right\}  -\frac{\sigma^{2}}{2N+1}\begin{cases}
                                                                            \frac{3N^2}{2} + \frac{5N}{2} + Nl + 1 - \frac{l}{2} - \frac{l^2}{2}, & 0\leq l  <N \\
                                                                            2N^2 + 3N + 1,                                                        & l\geq N
                                                                          \end{cases}
\end{align}
\newpage
\subsubsection*{Fourth part}
\begin{align}
    & \frac{1}{\left( 2N+1 \right)^{2}} \mathrm{E}\left[ \left( \sum\limits_{i=-N}^{N}\phi_{k+i} \right)\left( \sum\limits_{i=-N}^{N}\phi_{k+l+i} \right) \right]                                                                                                                                                                            \\
  = & \frac{1}{\left( 2N+1 \right)^{2}}\mathrm{E}\left[ \left(\left( 2N+1 \right)\sum\limits_{i=1}^{k-N-1}\Delta\phi_i + \sum\limits_{i=0}^{2N}\sum\limits_{j=0}^i \Delta\phi_{k-N+j}\right)\left(\left( 2N+1 \right)\sum\limits_{i=1}^{k+l-N-1}\Delta\phi_i + \sum\limits_{i=0}^{2N}\sum\limits_{j=0}^i \Delta\phi_{k+l-N+j}\right) \right] \\
  \begin{split}
    = & \frac{1}{\left( 2N+1 \right)^{2}}\Bigg[ \sigma^{2}\left( 2N+1 \right)^{2}\min\left\{ k-N-1,k+l-N-1 \right\}\\
      & + \left( 2N+1 \right)\mathrm{E}\left[ \left(\sum\limits_{i=1}^{k-N-1}\Delta\phi_i\right)\left( \sum\limits_{i=0}^{2N}\left( 2N+1 - i \right)\Delta\phi_{k+l-N+i} \right) \right]\\
      & + \left( 2N+1 \right)\mathrm{E}\left[ \left( \sum\limits_{i=0}^{2N}\left( 2N+1 - i \right)\Delta\phi_{k-N+i} \right)\left(\sum\limits_{i=1}^{k+l-N-1}\Delta\phi_i\right) \right]\\
      & + \mathrm{E}\left[ \left( \sum\limits_{i=0}^{2N}\left( 2N+1 - i \right)\Delta\phi_{k-N+i} \right)\left( \sum\limits_{i=0}^{2N}\left( 2N+1 - i \right)\Delta\phi_{k+l-N+i} \right) \right]
      \Bigg]
  \end{split}                                                                                                                                                                                                        \\
  \begin{split}
    = & \frac{1}{\left( 2N+1 \right)^{2}}\Bigg[ \sigma^{2}\left( 2N+1 \right)^{2}\min\left\{ k-N-1,k+l-N-1 \right\}\\
      & + 0\\
      & + \left( 2N+1 \right)\sigma^{2}\begin{cases}
        0,                                  & l < 1          \\
        2Nl - \frac{l^2}{2} + \frac{3l}{2}, & 1\leq l < 2N+1 \\
        2N^2 + 3N + 1,                      & l \geq 2N+1
      \end{cases}\\
      & + \sigma^{2}\begin{cases}
        \frac{8}{3}N^3 - 2N^2l + 6N^2 - 3Nl + \frac{13}{3}N - \frac{7}{6}l + 1 + \frac{l^3}{6}, & 0\leq l \leq 2N \\
        0,                                                                                      & l > 2N
      \end{cases}
      \Bigg]
  \end{split}                                                                                                                                                                                                        \\
  \begin{split}
    = & \frac{1}{\left( 2N+1 \right)^{2}}\Bigg[ \sigma^{2}\left( 2N+1 \right)^{2}\min\left\{ k-N-1,k+l-N-1 \right\}\\
      & + \sigma^{2}\begin{cases}
        \frac{8}{3}N^3 + 2N^2l + 2Nl -Nl^2 - \frac{l^2}{2} + 6N^2 + \frac{13}{3}N + 1 + \frac{l^3}{6} + \frac{1}{3}l, & 0\leq l \leq 2N \\
        4N^3 + 8N^2 + 5N + 1,                                                                                         & l > 2N
      \end{cases}\\
      &\Bigg]
  \end{split}
\end{align}
\subsubsection*{Combining all}
\renewcommand{\qedsymbol}{\ensuremath{\blacksquare}}
\begin{align}
  \begin{split}
    R_{\phi\phi}\left( l \right) = &
    \sigma^{2}\max\left\{ 0, N+1-l \right\}                                                                                                                                                                                                                                      \\
    & +\frac{\sigma^{2}}{\left( 2N+1 \right)^{2}}\begin{cases}
      \frac{\left|l\right|^3}{6} + l^2\left( N+\frac{1}{2} \right) + \left|l\right|\left( 2N^2 + 4N + \frac{4}{3} \right) - \frac{10}{3}N^3 - 7N^2 - \frac{14}{3}N - 1, & 0\leq \left|l\right| < N     \\
      \frac{\left|l\right|^3}{6} - l^2\left( N+\frac{1}{2} \right) + \left|l\right|\left( 2N^2 + 2N + \frac{2}{3} \right) - \frac{4}{3}N^3 - 2N^2 - \frac{2}{3}N ,      & N\leq \left|l\right| \leq 2N \\
      0,                                                                                                                                                                & \left|l\right| > 2N
    \end{cases}
  \end{split}
\end{align}

\end{document}